\documentclass[prd,aps,floats,twocolumn,nofootinbib]{revtex4-1}
\usepackage{slashed}
\usepackage{mathtools}
\usepackage{amsfonts}
\usepackage{amssymb}
\usepackage{epsfig}
\usepackage{bbold}

\begin{document}

\newcommand{\m}[1]{\mathcal{#1}}
\newcommand{\nn}{\nonumber}
\newcommand{\ph}{\phantom}
\newcommand{\eps}{\epsilon}
\newcommand{\be}{\begin{equation}}
\newcommand{\ee}{\end{equation}}
\newcommand{\bea}{\begin{eqnarray}}
\newcommand{\eea}{\end{eqnarray}}
\newtheorem{conj}{Conjecture}

\newcommand{\plk}{\mathfrak{h}}


\title{Unimodular Hartle-Hawking wave packets and their probability
    interpretation}
\date{ }


\author{Bruno Alexandre}
\author{Jo\~{a}o Magueijo}
\email{magueijo@ic.ac.uk}
\affiliation{Theoretical Physics Group, The Blackett Laboratory, Imperial College, Prince Consort Rd., London, SW7 2BZ, United Kingdom}

\begin{abstract}
We re-examine the Hartle-Hawking wave function from the point of view of a quantum theory which starts from the connection representation and allows for off-shell non-constancy of $\Lambda$ (as in unimodular theory), with a concomitant dual relational time variable. By translating its structures to the metric representation we find a non-trivial inner product rendering wave packets of Hartle-Hawking waves normalizable and the time evolution unitary; however, the implied  probability measure differs significantly from the naive $|\psi|^2$. In contrast with the (monochromatic) Hartle-Hawking wave function, these packets form travelling waves with a probability peak describing de Sitter space, except near the bounce, where the incident and reflected waves interfere, transiently recreating the usual standing wave. Away from the bounce the packets get sharper both in metric and connection space, an apparent contradiction with Heisenberg's principle allowed by the fact that the metric is not Hermitian, even though its eigenvalues are real. Near the bounce, the evanescent wave not only penetrates into the classically forbidden region but also extends into the $a^2<0$ Euclidean domain.  We work out the propagators for this theory and relate them to the standard ones. The $a=0$ point (aka the ``nothing'') is unremarkable, and in any case a wave function peaked therein is typically non-normalizable and/or implies a nonsensical probability for $\Lambda$ (which the Universe would preserve forever). Within this theory it makes more sense to adopt a Gaussian state in an appropriate function of $\Lambda$, and use the probability associated with the evanescent wave present near the time of the bounce as a measure of the likelihood of creation of a pair of time-symmetric semiclassical Universes. 
\end{abstract}

\maketitle

\section{Introduction}

The Hartle-Hawking wave function of the Universe was one of the first proposals of a concrete framework for quantum creation of the Universe out of nothing~\cite{HH}. Its interpretation and derivation has aroused much interest (e.g.~\cite{Jonathan,vil-rev,vil-PRD}), with a revival in recent years (e.g.~\cite{JeanLuc1,JeanLuc2,reply}). Even back in the 1980s, when the pioneering work of Hartle and Hawking was done, an almost orthogonal approach to quantum gravity was developed, making the connection, rather than the metric, the central character of the theory (e.g.~\cite{thiemann}). One of its earliest solutions was the  Chern-Simons-Kodama state~\cite{jackiw,kodama}, but in the face of its problems (e.g.~\cite{witten}) this was superseded by the loop representation, with sporadic backtracking~\cite{lee1,lee2,Randono0,Randono1,Randono2,Wieland,RealCS}.

The metric- and connection-driven  approaches then led separate lives. It was not until recently that it was realized that the Chern-Simons-Kodama state is in fact the Fourier dual of the Hartle-Hawking wave function under the most minimal assumption: that the connection is real~\cite{CSHHV}. 
The point of this paper is to explore what can be learnt from this duality regarding the initial conditions of the Universe.

The metric and connection appear as duals in the quantum theory and 
the choice of representation in quantum cosmology is not innocuous. It can lead to inequivalent theories: different natural ranges of variation for the variables and different natural inner products and probability interpretations, for example. It is only in the most standard setting, where the inner product is trivial and fixed a priori and the ranges for the variables are a given on physical grounds, that one may appeal to the Stone-von Neumann theorem and claim unitary equivalence between the representations. In contrast, in quantum cosmology the choice of representation may shed new light on the old issue of ``boundary conditions''. The reason why 
the position representation is usually favoured in standard Quantum Mechanics is that it is physically clearer for defining boundary conditions.  But in quantum
gravity/cosmology it is far from obvious which representation should receive primacy in this respect. In this paper we  reassess the metric driven no-boundary proposal and the Hartle-Hawking wave function from the point of view of a theory which starts from the connection representation. 

In order to do this, an extra ingredient is needed. The physical interpretation of the Chern-Simons-Kodama state is improved by a minimal extension of Einstein's gravity:  ``unimodular'' gravity~\cite{unimod1} in its fully diffeomorphism formulation~\cite{unimod}.  This happens for two reasons. First, the unimodular extension introduces a physical time variable (unimodular or 4-volume time~\cite{Bombelli}), so that the waves now move ``in physical time''. Second, it introduces a natural (unitary) inner product with respect to which normalizable wave packets, superposing states with different $\Lambda$, may be built~\cite{JoaoLetter,JoaoPaper}. From the point of view of unimodular theory, fixed-$\Lambda$ wave functions are just the ``spatial'' factors of monochromatic partial waves\footnote{``Spatial'' here is used not in the sense of $x$, which is trivial in minisuperspace, but in the sense of dependent on the non-time variables, here the metric or the connection, as well as the ``frequency'' conjugate to the time variable, here $\Lambda$ itself. This will be clearer later, cf. (\ref{psisb}) vs (\ref{HHpackets}), or (\ref{psisa2}) vs (\ref{HHpacketsa2}).}.  Pathologies found in the fixed-$\Lambda$ theory (infinite norms, lack of a time variable) are cured for the wave packets built from these partial waves, and physical behavior is found. In particular the peak of the probability (induced by the inner product) follows the classical trajectory in  unimodular time in the semi-classical regime~\cite{JoaoPaper}. Quantum deviations around this trajectory give predictability to the theory (see~\cite{bbounce,matLbounce} for examples). 

The narrative line of this paper is as follows. We first review results on the connection-driven (Section~\ref{sumary-connection}) and unimodular (Section~\ref{summary-unimodular}) quantum theories, as well as the metric-connection duality exposed in~\cite{CSHHV} (Section~\ref{summary-duals}). In Section~\ref{unimodmetric} we then translate into the metric formalism the construction of unimodular wave packets selecting the connection contour associated with Hartle-Hawking waves, paying particular attention to the inner product they acquire, the implied probability measure and the structure of the Hilbert space. 

Surprises are found at once. 
In Section~\ref{semiclassical} we explicitly construct packets of Hartle-Hawking waves for a Gaussian amplitude in $\phi=3/\Lambda$. These packets are dramatically different from the usual standing waves: they form well separated travelling waves even without the need of Vilenkin boundary conditions.  Their peaks follow the semiclassical limit and get sharper as the Universe gets larger, in an apparent contradiction with the Heisenberg principle which we resolve by inspection of the effects of our unusual inner product. 

In an attempt to make contact with creation out of nothing and the no-boundary proposal, we evaluate the unimodular propagators in Sections~\ref{unimodpropagators} and~\ref{retailor}. The mathematics is straightforward but serious problems are identified in Section~\ref{probsnoboundary} when trying to force the theory into creation out of $a=0$.  Briefly, such an initial condition is naturally non-normalizable, and would in any case imply a non-viable distribution for $\Lambda$ (which the Universe would have to live with for ever). Having started from the connection representation, the dual is also $a^2$, not $a$, with the whole real line naturally appearing in the theory.

All of this points to the creation of a pair of time-symmetrical semiclassical Universes out of the full evanescent wave resent around the bounce, with a Gaussian state, as argued in Section~\ref{outofwall}. In a concluding section we summarise the take-home messages of this paper.

\section{Summary of connection-based results}\label{sumary-connection}

The roots of connection led approaches, such as the Ashtekar formalism, are in the Einstein-Cartan (EC) formalism~\cite{thiemann}. We will only need the reduction to minisuperspace (MSS) in this paper, but start by presenting the full theory because this will illuminate some peculiarities.  
The EC action subject to a $3+1$ split 
takes the form:
\begin{eqnarray}
S_{EC}=\frac{1}{16\pi G }\int dt\, d^3x\, [2\dot{K}^i_aE^a_i-(NH+N^aH_a+N_iG^i)], \nonumber
\end{eqnarray}
where $K^i_a$ is the extrinsic curvature connection (from which the Ashtekar connection can be built by a canonical transformation~\cite{thiemann}), $E^a_i$ is the densitized inverse triad, and the last three terms are the Hamiltonian, Diffeomorphism and Gauss constraints, enforced by corresponding Lagrange multipliers. Quantization derives from implementations of:
\begin{eqnarray}\label{commAsh}
[K^i_a(\boldsymbol{x}),E_j^b(\boldsymbol{y})]=il_P^2\delta^b_a\delta^i_j\delta(\boldsymbol{x}-\boldsymbol{y})
\end{eqnarray}
where $l_P=\sqrt{8\pi G \hbar }$ is the reduced Planck length.

Adding $\Lambda$, in MSS this action becomes (e.g.~\cite{CSHHV,MZ}):
\begin{equation}\label{Sg}
S_0=\frac{3V_c}{8\pi G} \int 
dt\bigg(\dot{b}a^2-Na \left[- (b^{2}+k 
)+ \frac{\Lambda}{3}a^2\right]\bigg),
\end{equation}
where $a$ is the expansion factor, $b$ is the only MSS connection variable (an off-shell version of the Hubble parameter, since $b= \dot a$ on-shell, if there is no torsion), $k$ is the normalized spatial curvature (assumed $k=1$, as usual), $N$ is the lapse function and $V_c=\int d^3 x$ is the comoving volume of the region under study, assumed finite throughout this paper (in the quantum cosmology classical literature one usually chooses $k=1$ and $V_c=2\pi^2$; see~\cite{Afshordi} for a discussion of the criteria for the choice of $V_c$). Hence, (\ref{commAsh}) becomes:
\begin{equation} \label{com1}
\left[ \hat{b},\hat{a^{2}}\right] =i\frac{l_{P}^{2}}{3V_{c}}\equiv i\plk, 
\end{equation}
so that in the $b$ representation:
\be\label{a2inb}
\hat a^2=-i\frac{l_P^2}{3V_c}\frac{\partial}{\partial b}=-i\plk \frac{\partial}{\partial b},
\ee
leading (with suitable ordering) to the WDW equation:
\be\label{WDW}
\left[-(b^{2}+k)-i\plk  \frac{\Lambda}{3}\frac{\partial}{\partial b}\right]\psi_s=0,
\ee
where we introduce the subscript $s$ to solutions of the WDW equation for later conveninece. 
This is solved by the Chern-Simons-Kodama (CSK) state reduced to MSS:
\begin{equation}\label{psisb}
    \psi_s(b,\phi)=\psi_{CS}(b,\phi)={\cal N}_b\exp{\left[\frac{i}{\plk } \phi X (b) \right]},
\end{equation}
where ${\cal N}_b$ is a normalization factor\footnote{Irrelevant in much of the original work on the CSK state, but essential here, as we will see.}, $\phi=3/\Lambda$ and 
\begin{equation}
    X(b)={\cal L}_{CS}=\frac{b^3}{3}+k b
\end{equation}
is the MSS reduction of the Chern-Simons functional.
It is known~\cite{CSHHV} that the CSK state is the Fourier dual of both the Hartle-Hawking (HH) and Vilenkin (V)wave functions, depending on the choice of contour (and sign of $\Lambda$, as we will see). Indeed the literature on the HH and V state uses the CSK state unwittingly (see~\cite{Jonathan} for example).  Although not strictly needed {\it in this paper we assume that $b$ covers the whole real line} (so that we have a HH dual for $\Lambda>0$). Dropping this assumption will be investigated elsewhere: there are some technical differences. 


As announced in the introduction, the representation from which one starts matters. By starting from the connection we have made the following choices which are not innocuous when re-examined from the metric viewpoint:
\begin{itemize}
    \item The natural variable is $\phi=3/\Lambda$ and not $\Lambda$. Classically this amounts to a canonical transformation: $\Lambda \rightarrow \phi(\Lambda)$ and $T\rightarrow T_\phi=T/\phi'(\Lambda)$. The quantum mechanical theories that follow are not equivalent, as we shall presently see. 
    
     \item Starting from the metric we are led to the pair $\{a,p_a\}$ (possibly with $a>0$), whereas starting from the connection the natural pair is $\{b,a^2\}$. This is because the conjugate of the connection is the densitized inverse triad $E^i_a$, and in MSS this is $a^2$. Again they are canonically related, but lead to quantum theories naturally based on different assumptions. Instead of $a>0$, in the $b$ representation $a^2$ should cover the whole real line including the negative-Euclidean section. This is because $b$ generates translations in $a^2$, and remains most naturally Hermitian if left to act unencumbered. 
\end{itemize}


\section{Review of the unimodular Chern-Simons state}
\label{summary-unimodular}

We use the Henneaux and Teitelboim formulation of ``unimodular'' gravity~\cite{unimod}, where full diffeomorphism invariance is preserved (so that ``unimodular'' is actually a misnomer). In this formulation one adds to $S_0$  a new term:
\be\label{Utrick}
S_0\rightarrow S=
S_0- \frac{3}{8\pi G }  \int d^4 x \, \phi\,  \partial_\mu T^\mu
\ee
(the pre-factor is chosen for later convenience). Here $T^\mu$ is a  density, so that the added term is diffeomorphism invariant without the need of a $\sqrt{-g}$ factor in the volume element or of the connection in the covariant derivative. Since the metric and connection do not appear in the new term, the Einstein equations and other standard 
field equations are left unchanged. 
The only new equations of motion are:
\begin{eqnarray}
\frac{\delta S}{\delta T ^\mu}=0&\implies& \partial_\mu \phi=\partial_\mu \Lambda =0\\
\frac{\delta S}{\delta \Lambda }=0&\implies& \partial_\mu  T^\mu\propto \sqrt{-g}
\end{eqnarray}
i.e. on-shell-only constancy for $\Lambda$ (the defining characteristic of unimodular theories~\cite{unimod1,unimod,alan,daughton,sorkin1,sorkin2}) and  the fact that $T^0$ is proportional to a prime candidate for relational time: 4-volume time~\cite{unimod,unimod1,UnimodLee1,Bombelli,UnimodLee2}. 

Reduction to MSS gives:
\be
S_0\rightarrow S=
S_0 +  \frac{3 V_c}{8\pi G }  \int dt x \, \dot \phi\,  T
\ee
(where we identify $T\equiv T^0$), so classically nothing changes except that we gain a canonical pair enforcing the constancy of $\Lambda$ as an equation of motion, and a ``time'' variable:
\begin{equation}\label{timeformula}
\dot T =N\frac{ a^3}{\phi^2}=N\frac{\Lambda^2}{9}a^3.
\end{equation}
However, the quantum mechanics is very different, since:
\bea\label{com2}
\left[\phi,T\right]&=&i\plk,
\eea
that is $\phi$ and $T$ are quantum complementaries. Hence, we can choose either the $\phi$ (i.e. $\Lambda$) representation, leading to the WDW equation (\ref{WDW}) for $\psi_s(b,\phi)$, or its dual time representation, leading to a Schrodinger equation:
\be\label{WDWSchro1}
\left[-i\plk \frac{1}{b^2+k}\frac{\partial}{\partial b}- i \plk \frac{\partial }{\partial T}\right]\psi(b,T) =0,
\ee
for a wave function depending on time $T$ instead. From the unimodular perspective~\cite{CSunimod} the CSK state 
is just the spatial factor, 
$\psi_s=\psi_{CS}$, of a monochromatic wave (with fixed $\Lambda$) moving in unimodular time $T$ conjugate to $\phi$. The  general solution to the Hamiltonian constraint is the superposition:
\bea
\psi(b,T)&=&
\int^\infty_{-\infty} 
d\phi 
{\cal A}(\phi) \exp{\left[-\frac{i}{\plk } \phi T \right]}\psi_s(b,\phi),
\nn\\
&=&\int^\infty_{-\infty}  \frac{d\phi}{\sqrt{2\pi\plk}} {\cal A}(\phi) \exp{\left[\frac{i}{\plk } \phi (X (b)-T ) \right]},\label{HHpackets}
\eea
for some amplitude function ${\cal A}(\phi)$. We have chosen normalization ${\cal N}_b^2=|\psi_s|^2=1/ (2\pi\plk)$ so that the inversion formula is symmetric:
\begin{equation}
{\cal A}(\phi) =
\int   dX\, \psi(b,T) \frac{e^{-\frac{i}{\plk } \phi (X-T)}}{\sqrt{2\pi\plk}}.
\label{Ampb}
\end{equation}
For a Gaussian amplitude centered on $\phi_0$ leading to a probability with variance $\sigma_\phi$ we find wave packets:
\bea\label{Gausspacketsb}
\psi (b,T)&=& \frac{e^{-\frac{i}{\plk } \phi_0 (X-T)}}{(2\pi\sigma_T^2)^\frac{1}{4}}\exp\left[-\frac{(X -T )^2}{4\sigma_T^2}\right]
\eea
with 
\begin{equation}\label{sigmaT}
   \sigma_T=\frac{\plk}{2\sigma_\phi} 
\end{equation}
saturating the Heisenberg uncertainty relation following from (\ref{com2}).


Within the unimodular perspective, 
the natural inner product between two states is given by:
\be\label{innalpha}
\langle\psi_1|\psi_2  \rangle=\int d\phi \,  {\cal A}_1^\star (\phi) {\cal A}_2(\phi) ,
\ee
and this product is automatically conserved with respect to time $T$, i.e. unitarity is enforced, since it is defined in terms of $T$-independent amplitudes. By virtue of Parseval's theorem
(with the assumption that $b$ is real) this product is 
equivalent:
\be\label{innX}
\langle\psi_1|\psi_2  \rangle=\int dX  \psi_1^\star(b,T)\psi_2(b,T).
\ee
Hence the probability in terms of $b$ is:
\be
{\cal P}(b)=|\psi(b,T)|^2\frac{dX}{db}=|\psi(b,T)|^2(b^2+k)
\ee
where we note the measure factor\footnote{This inner product is exact and should not be confused with its approximate  semi-classical cousins required in more complicated situations (such as multi-fluids or minority clocks~\cite{JoaoPaper,bbounce,bruno}).}.
Note that this could have been guessed directly from the conserved current:
\be\label{current}
j^T=j^X=|\psi|^2
\ee
associated with the Schrodinger equation (\ref{WDWSchro1}) written as:
\be
\left(\frac{\partial}{\partial X}+
 \frac{\partial}{\partial T}
\right)\psi =0,
\ee
($\partial_a j^a=0$, for $a=T,X$). Note also that one can bypass expansion (\ref{HHpackets}) to find the the general solution:
\be\label{FTXalpha}
\psi(b,T)=F(T -X),
\ee
where $F$ can be any function. Hence the waves written in terms of $X$ are non-dispersive. This allows us to guess the propagator directly. 

Given that we have unitarity, it is reasonable to define physical states as any state derived from an amplitude ${\cal A}(\phi)$ such that:
\be\label{normalpha}
\langle\psi|\psi  \rangle=\int d\phi \,  |{\cal A} (\phi)|^2 =1,
\ee
that is any state with norm 1. Hence a delta function is not acceptable since the integral of the square of a delta function is not 1, supporting the view aired previously that fixed $\Lambda$ ``monochromatic'' waves (whether HH or V, or CSK states) are not physical, because non-normalizable. The Gaussian states are normalizable, and as we will see, lead to a sound semi-classical limit. However, we stress that not all ${\cal A}(\phi)$ lead to a semiclassical limit, so there is no a priori reason why perfectly physical states should be semi-classical at all. For example, a normalized uniform distribution in $\phi$ (confined within a finite range) is never semi-classical. We should not be surprised that semi-classicality is a matter of choice/selection of state, rather than an imposition from mathematical consistency, or physical Hilbert space. Fully and endemic quantum behaviour is perfectly physical.

We close this review by specifying why the canonical transformations $\Lambda \rightarrow \phi(\Lambda)$ and $T\rightarrow T_\phi=T/\phi'(\Lambda)$, lead to 
theories which are all classically equivalent (and in fact have the same semiclassical limit), but their quantum mechanics is different. 
Their solutions (\ref{HHpackets}) are different: a Gaussian in $\Lambda$ is not a Gaussian in a generic $\phi(\Lambda)$; the frequency $\Lambda T$ is not invariant under the canonical transformation. The natural unimodular inner product (\ref{innalpha}) is also not invariant~\cite{bbounce,JoaoPaper}. Although all these quantum theories are different, for a generic $\phi$ chosen {\it within reason}, their border with the semi-classical limit is the same, as we will comment in more detail later.

\section{The monochromatic metric duals of the CSK state}\label{summary-duals} 
Our first purpose is to translate the constructions reviewed in the last two Sections to the metric representation. We start with the metric dual of the CSK state, extending~\cite{CSHHV} to suit our purposes. We assume that both $b$ and $a^2$ are real and unconstrained, so that:
\begin{eqnarray}
\psi_s(a^2,\phi)&=&\int^\infty_{-\infty}\frac{db}{\sqrt{2\pi\plk}}e^{-\frac{iba^2}{\plk}}\psi_s(b,\phi)\label{FTba2}\\
\psi_s(b,\phi)&=&\int^\infty_{-\infty}\frac{da^2}{\sqrt{2\pi\plk}}e^{\frac{iba^2}{\plk}}\psi_s(a^2,\phi)\label{FTa2b}
\end{eqnarray}
(in~\cite{CSHHV} we did not define (\ref{FTa2b}); note the symmetrical convention adopted here in contrast with~\cite{CSHHV}). Note that (\ref{normalpha}) implies boundary conditions both as $a^2\rightarrow\infty$ and $a^2\rightarrow\infty $. 
We will often assume that ${\cal A}(\phi)$ is peaked at a positive $\phi$ and exponentially suppressed at $\phi<0$, but in some parts of this paper a general $\phi$ will be required. It was 
shown in~\cite{CSHHV} that the CSK state is the Fourier dual of both the HH and V wave functions, depending on the choice of contour and that imposing the reality of $b$ selects the HH wave function if $\Lambda>0$ (which for simplicity and definiteness was assumed throughout~\cite{CSHHV}). We will relax this assumption here, and not only confirm the results of~\cite{CSHHV} for $\Lambda>0$, but also show that the reality of $b$ selects the V wave function if $\Lambda<0$. 

This is straightforward to show using the integral representation of the Airy-like functions:
\begin{eqnarray}\label{intrepAi}
f(-z)=\frac{1}{2\pi}\int^\infty_{-\infty} e^{i\left(\frac{t^3}{3}-zt\right)}dt,
\end{eqnarray}
noting that it maps onto (\ref{FTba2}) with~\cite{CSHHV}:
\begin{eqnarray}
&& z=-\left(\frac{\phi}{\plk}\right)^{2/3}
\left(k-\frac{ a^2}{\phi}\right), \label{zdef}\\
&& t=\left(\frac{\phi}{\plk}\right)^{1/3}b.
\end{eqnarray}
For $\phi>0$, a real $b$ implies a real $t$, so that (\ref{intrepAi}) gives an Airy function:
\begin{eqnarray}\label{psisa2}
\psi_s\left(a^{2},\phi\right)&=&
{\cal N}_a {\rm Ai} (-z)\label{psisAi}\\
{\cal N}_a&\equiv& \frac{1}{\phi^{1 / 3} \plk^{2 / 3}}
\end{eqnarray}
that is, the HH wave function suitably normalized to match the normalization ${\cal N}_b$ chosen for its connection dual\footnote{We notice that this normalization {\it is at odds with} the assertion made in~\cite{vil-PRD} that $\psi_s(0)$ should be a constant as a function of $\Lambda$ (see their Eq.~4.19), an assumption that ultimately pegs down the final result for the probability of tunneling in~\cite{vil-PRD}. The apparent contradiction results from the fact that the two theories are different (the one in~\cite{vil-PRD} identifies $\Lambda$ with the potential energy domination of a scalar field; the $\Lambda$ here is an unimodular $\Lambda$).}. For $\phi<0$, however, a real $b$ implies a contour in $t$ shifted from the real axis by $e^{i\pi/3}$, producing the V wave function:
\begin{eqnarray}\label{psisaibi}
\psi_s\left(a^{2},\phi \right)&=&
\frac{{\cal N}_a}{2} [{\rm Ai} (-z)+ i{\rm Bi} (-z)]
\end{eqnarray}
(where the modulus of $\phi$ is to be used in the first factor in (\ref{zdef}) so that the argument is real).
This is not surprising, since the HH and V wave functions are known to be related by $\phi\rightarrow -\phi$, $a^2\rightarrow -a^2$ (or
$\Lambda\rightarrow e^{i\pi} \Lambda$, $a\rightarrow e^{i\pi/2} a$, as~\cite{vil-PRD} puts it). 
Notice that given that $k=1$ and $\Lambda<0$ there are no Lorentzian classical solutions  (cf. the Hamiltonian constraint). All the solutions will necessarily be Euclidean, indeed related to the ones we are about to find via a $a^2\rightarrow -a^2$ transformation.

This detail will be largely a formality, since we will mostly choose amplitudes sharply peaked around a positive $\Lambda$. However, in generic formal calculations we should not forget that the basis functions $\psi_s(a^2,\phi)$, usually the real HH (stationary) wave functions for $\Lambda>0$, are not generally real within the unimodular formalism, since a priori we cannot assume that $\cal A$ only has support on $\phi>0$. This clears apparent contradictions when later in this paper we evaluate the propagators (Section~\ref{unimodpropagators}, in particular around (\ref{Greena2}) and footnote 8). These apparent contradictions only arise if we forget this detail.


\section{Unimodular metric theory}
\label{unimodmetric}
Moving on to the unimodular theory, we can define the metric-representation superpositions as:
\bea
\psi(a^2,T)&=&
\int^\infty_{-\infty} 
d\phi\, 
{\cal A}(\phi) \exp{\left[-\frac{i}{\plk } \phi T \right]}\psi_s(a^2,\phi)\label{HHpacketsa2}
\eea
so that we have:
\begin{eqnarray}
\psi(a^2,T)&=&\int\frac{db}{\sqrt{2\pi\plk}}e^{-\frac{iba^2}{\plk}}\psi(b,T)\label{FTpack-ab}\\
\psi(b,T)&=&\int\frac{da^2}{\sqrt{2\pi\plk}}e^{\frac{iba^2}{\plk}}\psi(a^2,T),\label{FTpack-ba}
\end{eqnarray}
mimicking (\ref{FTba2}) and (\ref{FTa2b}) for the fixed-$\Lambda$ theory. 
This is true because the integrals in $\phi$ and those in $b$ and $a^2$ commute. 
We can also introduce the $\phi$ eigenstates as seen by unimodular theory
\begin{equation}
    \psi_\phi(q,T)=\psi_s(q;\phi)e^{-\frac{i}{\plk}\phi T},
\end{equation} 
with $q=b,a^2$ for the metric and connection representations, respectively. These are the full expressions (with time factors) for the monochromatic partial waves, and we can write the packets as: \bea
\psi(q,T)&=&
\int^\infty_{-\infty} 
d\phi\, 
{\cal A}(\phi)\psi_\phi (q,T), \label{HHPacketsq}
\eea
in lieu of (\ref{HHpackets}) and (\ref{HHpacketsa2}).


\subsection{Metric representation of the inner product}
We can now translate into the metric representation the inner product that most naturally arises in the connection representation of unimodular theory (i.e. (\ref{innalpha}) or (\ref{innX})). 
Substituting (\ref{FTpack-ba}) in (\ref{innX}) we find:
\begin{eqnarray}
\langle\psi_1|\psi_2\rangle=\int db|\partial_b X|\frac{da^2da'^2}{2\pi\plk}e^{\frac{ib}{\plk}(a'^2-a^2)}\psi_1^*(a^2,T)\psi_2(a'^2,T).\nn 
\end{eqnarray}
Given that $\partial_b X=b^2+k$, we can
use:
\begin{eqnarray}
\int \frac{db}{2\pi\plk}e^{\frac{i}{\plk}b\Delta a^2}=\delta(\Delta a^2),
\end{eqnarray}
\begin{eqnarray}
\int dbb^2e^{\frac{i}{\plk}b\Delta a^2}=-2\pi\plk^3\delta''(\Delta a^2)
\end{eqnarray}
(where $\Delta a^2 =a'^2-a^2$) to obtain, after two integrations by parts:
\begin{eqnarray}\label{inna2}
\langle\psi_1|\psi_2\rangle&=&\int^\infty_{-\infty} da^2\big(k\psi_1^\star (a^2,T)\psi_2 (a^2,T)
\nn\\
&&+\plk^2  \partial_{a^2} \psi_1^\star (a^2,T)\partial_{a^2} \psi_2 (a^2,T)\big).
\end{eqnarray}
This is the general form of the metric representation of our inner product\footnote{To the best of our knowledge this inner product is different from any previously proposed in the literature.}, and given that it ultimately comes from (\ref{innalpha}), its time-independence is guaranteed. If one of the states is an eigenstate of $\phi$ (with whatever normalization) we can perform and integration by parts and use the Airy equation to simplify its expression to:
\begin{eqnarray}\label{inna2b}
\langle\psi_1|\psi_2\rangle&=&\int^\infty_{-\infty}
da^2\, \frac{a^2}{\phi} \psi_1^\star (a^2,T)\psi_2 (a^2,T)\nn\\
&=&\int^\infty_{-\infty}
d\mu(a^2) \psi_1^\star (a^2,T)\psi_2 (a^2,T).
\end{eqnarray}
{\it In this case only}, the metric inner product is a change in the the measure factor, $d\mu(a^2)$, which happens to be identical do $dX(b)$ using the Hamiltonian constraint for fixed $\phi$ (since $dX=(b^2+k)db$ and $b^2+k=a^2/\phi$). 

Later on we will need
metric representation counterpart to inversion formula (\ref{Ampb}).
Inserting (\ref{FTpack-ba}) into (\ref{Ampb}) leads to:
  \begin{eqnarray}\label{Ampa^2}
    {\cal A}(\phi) &=& \int  da^2
    (k \psi_\phi^\star  (a^2,T)
    \psi (a^2,T)
\nn\\
&&+\plk^2  \partial_{a^2}  \psi^\star _\phi (a^2,T) \partial_{a^2} \psi (a^2,T)\big) 
\end{eqnarray}
mimicking the inner product (\ref{inna2}). This is hardly surprising, with (\ref{Ampb}) and  (\ref{Ampa^2}) expressing: 
\begin{equation}\label{ampbraket}
    {\cal A}(\phi) =\langle \phi|\psi \rangle
\end{equation}
where we denote $|\phi\rangle$ the $\phi$ eigenstate, in whatever representation. We can also apply the simplification (\ref{inna2b}), leading to:
\begin{eqnarray}\label{ampa2b}
    {\cal A}(\phi) &=&\int  d\mu(a^2)\, \psi^\star_\phi  (a^2,T) \psi (a^2,T).
\end{eqnarray}

\subsection{The structure of the wave functions space}
Once we recognize the non-trivial nature of the inner products in both representations some of our results can be put into the standard bra and ket notation. The unimodular general solutions (\ref{HHpackets}) and (\ref{HHpacketsa2}) can be seen as an expression of the partition:
\be\label{1phi}
\mathbb{1}=\int d\phi \, |\phi\rangle \langle \phi |
\ee
with:
\begin{eqnarray}
\psi(q,T)&=&\langle q T| \psi\rangle\\
{\cal A}(\phi) &=& \langle \phi | \psi\rangle\\
\psi_\phi (q,T)&=& \langle q T| \phi\rangle \\
\psi_s(q;\phi)&=&\langle q | \phi\rangle
\end{eqnarray}
with factorization for the eigenfunctions:
\begin{eqnarray}
\langle q T| \phi\rangle =\langle q | \phi\rangle \langle T| \phi\rangle = \psi_s(q;\phi)e^{-\frac{i}{\plk}\phi T}.
\end{eqnarray}
That is, we can rewrite our general solutions in the equivalent form:
\bea\label{factorbracket}
\langle q T| \psi\rangle &=& 
\langle q T| \left (\int d\phi \, |\phi\rangle \langle \phi| \right)|\psi\rangle 
\nn\\
&=&
\int d\phi\,  \langle \phi | \psi\rangle \langle q T| \phi\rangle \nn\\
&=& \int d\phi\,  \langle \phi | \psi\rangle \langle q | \phi\rangle \langle T| \phi\rangle \nn\\
\Leftrightarrow \quad  \psi(q,T)&=&\int
d\phi\, 
{\cal A}(\phi)\psi_\phi (q,T),\nn\\
&=&
\int  
d\phi\, 
{\cal A}(\phi) \exp{\left[-\frac{i}{\plk } \phi T \right]}\psi_s(q,\phi)
\eea
Notice also how the form of the inner product (\ref{innalpha}) can be obtained by an insertion of (\ref{1phi}), given (\ref{ampbraket}). 


A similar formal development can be applied for $q=\{b,a^2\}$ with a proviso. We would want the inner products (\ref{innX}) and (\ref{inna2}) to result from an insertion of a completeness relation, but clearly we would need a special $\otimes$ reflecting the non-trivial inner product (\ref{inna2}):
\bea
\mathbb{1}=\int dq \, |q\rangle\otimes  \langle q |
\eea
with:
\begin{equation}
    |a^2\rangle \otimes   \langle a^2 |= 
    k|a^2\rangle \langle a^2 |+
    \plk^2|a^2\rangle\overleftarrow{\partial_{a^2}}  \overrightarrow{\partial_{a^2}} \langle a^2 |
\end{equation}
in the case of $a^2$. 
We do know that if this is inserted into at least one eigenstate of $\phi$ the expression simplifies. In such as case, and for $q=b$ the completness relations are just:
\bea
\mathbb{1}=\int d\mu(q) \, |q\rangle  \langle q |
\eea
Expressions (\ref{Ampb}) and (\ref{ampa2b}) are such insertions into (\ref{ampbraket}).



\subsection{Probability interpretation}
The inner product (\ref{inna2}) implies the exact unitary definition of probability in metric space (with measure $da^2$):
\begin{eqnarray}\label{proba2}
{\cal P}(a^2) = k|\psi(a^2,T)|^2+\plk^2|\partial_{a^2}\psi(a^2,T)|^2,
\end{eqnarray}
and
\begin{equation}
    \int da^2\,  {\cal P}(a^2,T)=1
\end{equation}
is guaranteed by (\ref{innalpha}) (and has been extensively verified numerically for all the solutions shown in this paper, as a  check). 
This probability density is very interesting. It contains a term controlled by the curvature $k=1$, which is just the Born probability, using $d a^2$ as a measure. It then contains a new term related to the derivative of the wave function, controlled by pre-factor $\plk$. Contrary to what might be expected, as we shall see in the next Section, in the semi-classical limit the {\it second} term dominates:
\begin{eqnarray}\label{proba2approx}
{\cal P}(a^2) \approx  \plk^2|\partial_{a^2}\psi(a^2,T)|^2,
\end{eqnarray}
with the more obvious first term  only relevant at the time of the bounce ($T\sim 0$) when the incident and reflected wave interfere, as well as in the classically forbidden region. 

This probability not only integrates to 1 at all times but in the relevant case $k=1$ is positive definite. The same happens if $k=0$, but not if $k=-1$. In this case, our definition shares with the Wigner function and some other quasi-probabilities, the fact that in strongly non-classical situations it is not positive definite. Indeed it would be interesting to work out the Wigner functions for these theories. Note that we are not implying that $k=-1$ should be discounted as unphysical because of this; merely that the probability interpretation for this $k$ and deep in the quantum phase (where the $k$ term is relevant in
(\ref{proba2})) is more subtle than defining a conserved measure. As with the Wigner function, the density providing the norm still integrates to 1, but it is not positive definite. This oddity complies with the correspondence principle as laid down in~\cite{Vilenkin-interpretation}, with the difference that we require less drastic novelties (such as loss of unitarity) deep in the Planck epoch.



\section{Hartle-Hawking wave packets and the semi-classical limit}\label{semiclassical}

\begin{figure}[h!]
    \centering
\includegraphics[scale=0.53]{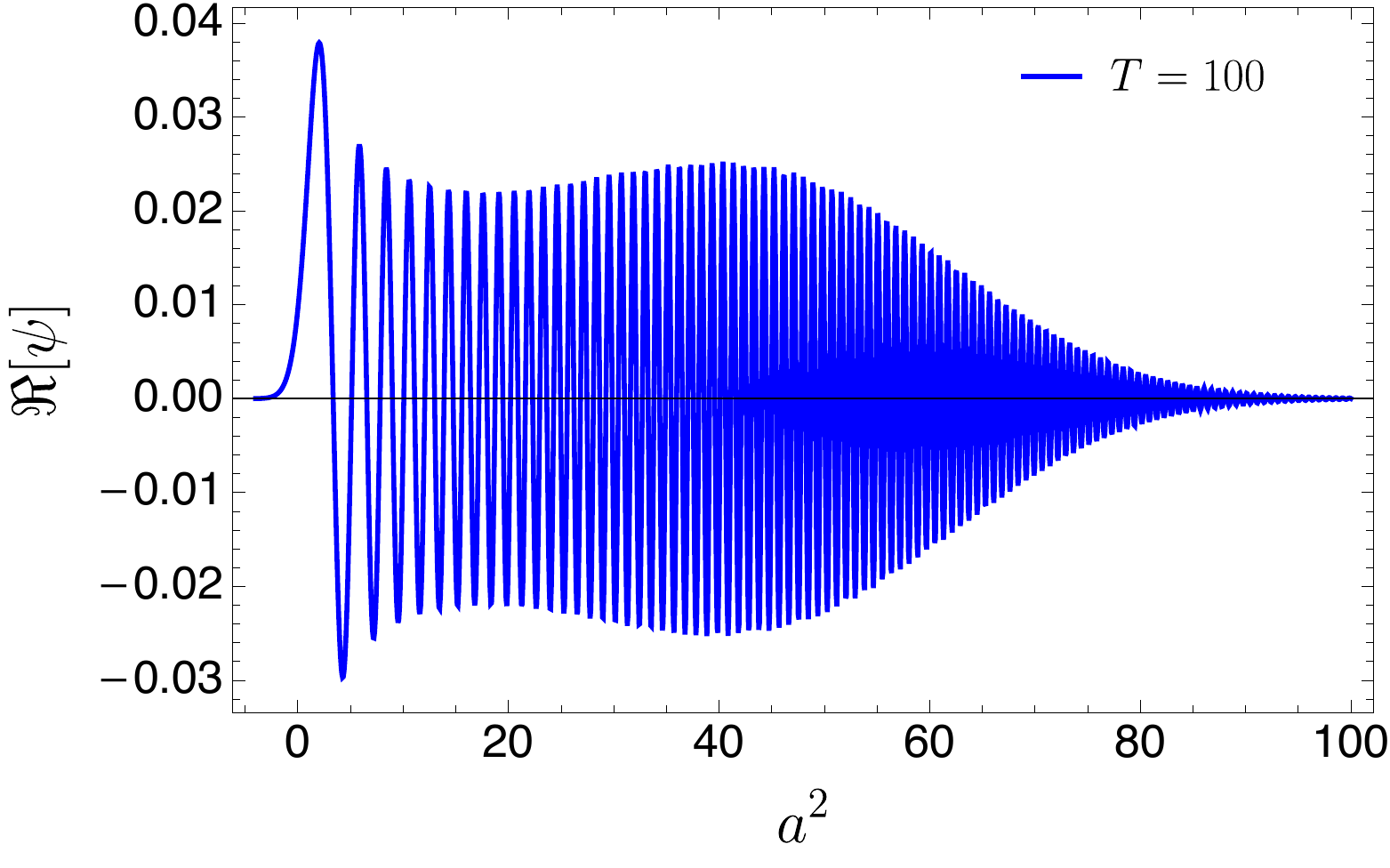}
\includegraphics[scale=0.53]{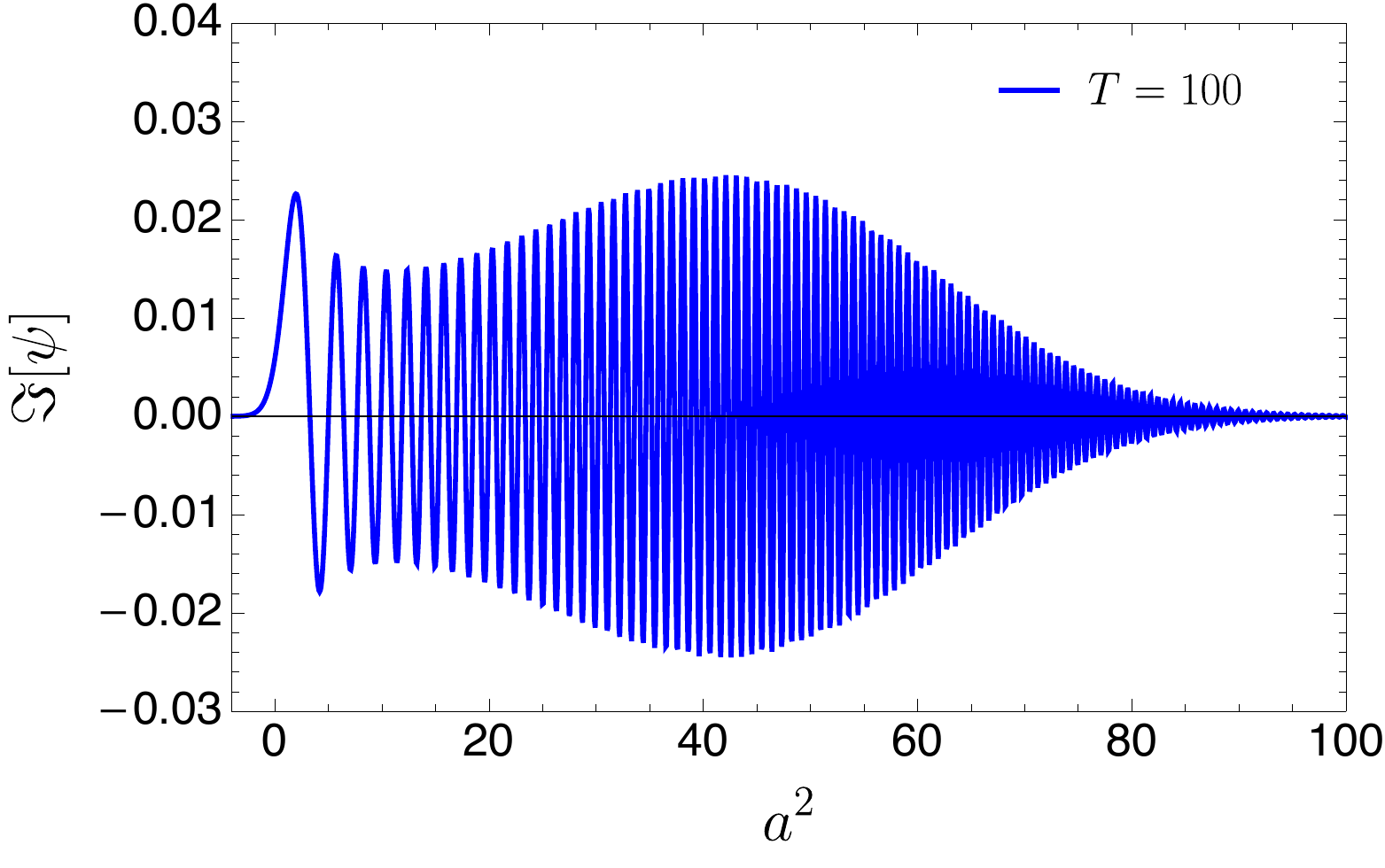}
\caption{Real (top panel) and imaginary (bottom panel) parts of the wave packet  $\psi(a^2,T)$ as a function of $a^2$ at $T=100$, for $\phi_0=1$, $k=1$, $\sigma_{\phi}=0.01$ (and $\plk=1$ implying a $\sigma_T=50$ according to the saturated Heisenberg relation (\ref{sigmaT})).} 
\label{figapsi100}
\end{figure} 

 \begin{figure}[h!]
\centering
\includegraphics[scale=0.53]{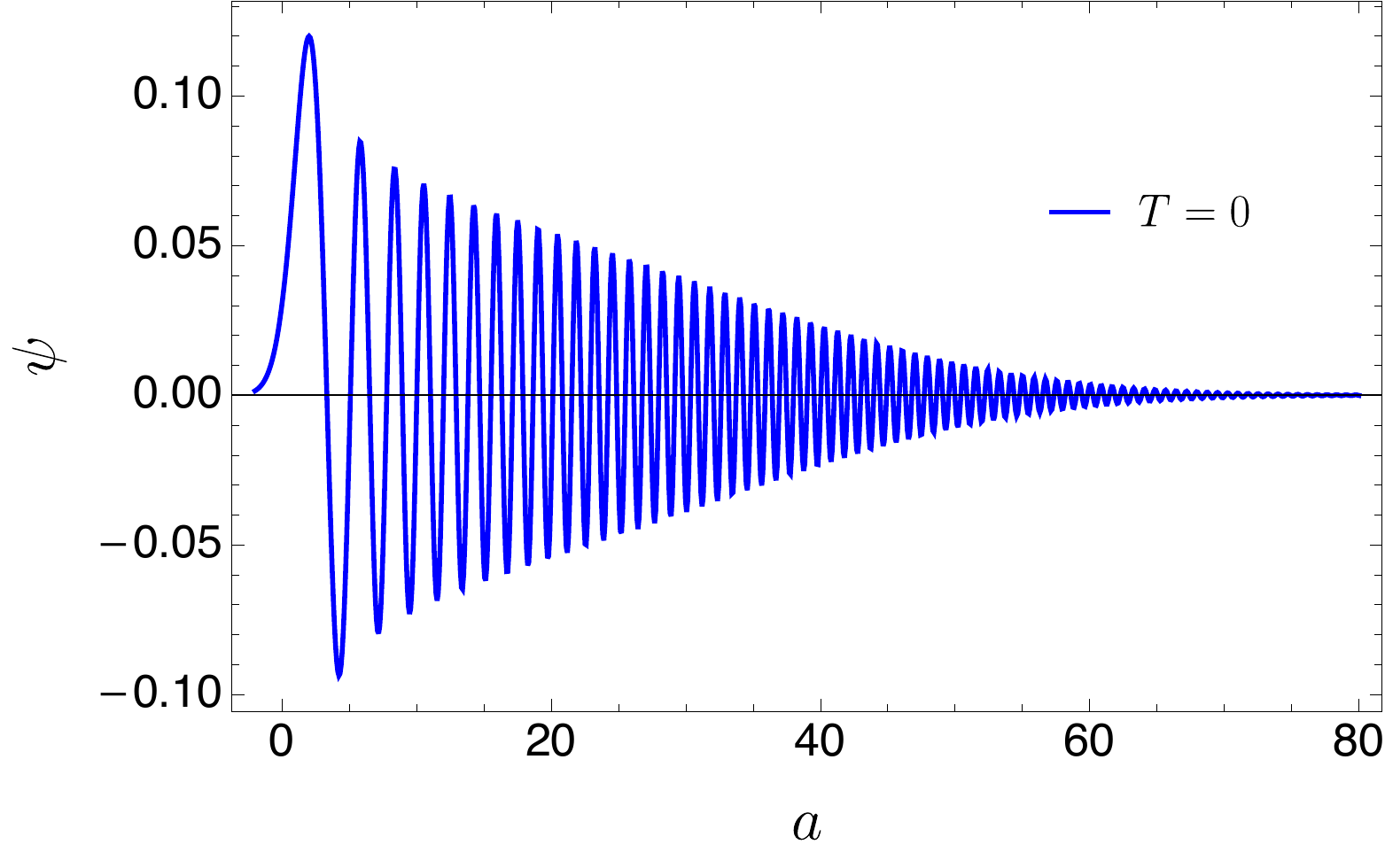}
\includegraphics[scale=0.53]{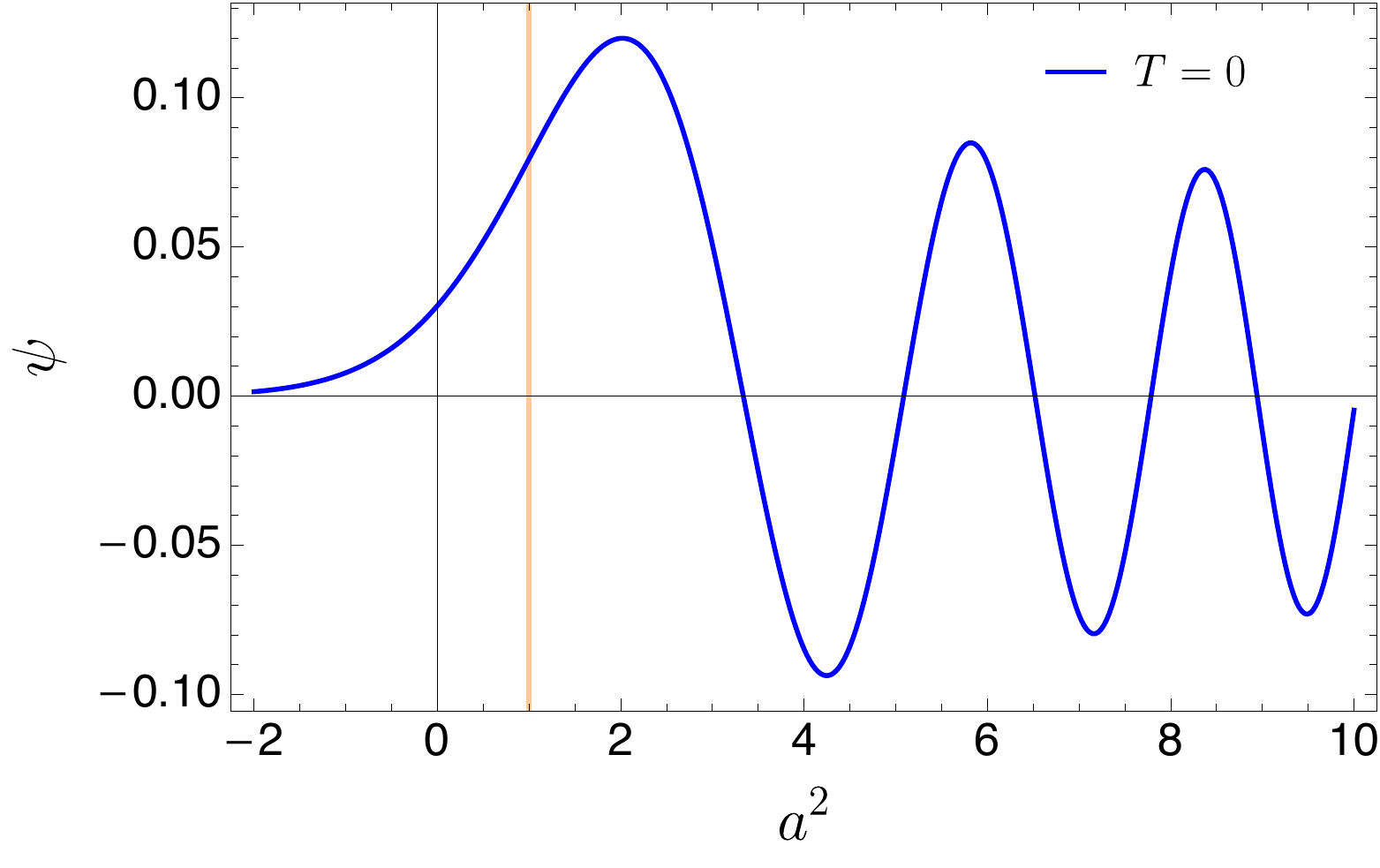}
\caption{The same wave packet $\psi(a^2,T)$ at $T=0$. 
The bottom panel is a close up of the top panel.  At time $T=0$ the wave function is real, forming an instantaneous standing wave resulting from the superposition of perfectly symmetric incident and reflected waves.   
The bottom panel close up illustrates the evanescent wave penetrating the classically forbidden region and smoothly extending to the Euclidean section, $a^2<0$ (the orange vertical line represents $a=a_*$). There is nothing special about the point $a^2=0$ corresponding to the ``no-boundary proposal''.  }
\label{figapsiT0}
\end{figure}

We can now build wave packets by inserting the amplitude:
\begin{equation}\label{Gaussamp}
 {\cal A}(\phi) =\sqrt{   N(\phi;\phi_0,\sigma_{\phi})}
\end{equation}
(where $N$ denotes a normal distribution)
into (\ref{HHpacketsa2}). For this choice, 
the distribution of $\phi$ (associated with inner product (\ref{innalpha})) is a  Gaussian centred at $\phi_0>0$ with standard deviation $\sigma_{\phi}$, with $\sigma_{\phi}\ll\phi_0$ assumed. 
Note that the integral (\ref{HHpacketsa2}) is then negligible for $\phi< 0$, hence we have essentially packets of HH wave functions (i.e. with $\psi_s$ given by (\ref{psisAi}) and not (\ref{psisaibi})). We have numerically checked this fact. 
Recall that this amplitude implies an uncertainty in time $\sigma_T$ according to the saturated Heisenberg relation (\ref{sigmaT}).
Illustrative examples of the numerical integration of (\ref{HHpacketsa2}) are plotted in Figs.~\ref{figapsi100}-\ref{figapsiProbT0}. The salient features are:

\begin{itemize}
    \item Unlike for the HH wave function, its corresponding unimodular packets are not generally real (even when the amplitude ${\cal A}$ is real), because the time factor of each partial wave is complex (see Eq.(\ref{HHpacketsa2})). Consequently the unimodular packets are not standing waves, even though the $\psi_s$ are standing waves (for $\Lambda>0$, as assumed here). See Fig.~\ref{figapsi100}. 
    \item  
    In contrast with the HH wave function, these packets are localized in $a^2$, i.e. have exponential fall-off on either side of their envelope's peak (see Fig.~\ref{figapsi100} and~\ref{figapsiT0} for an illustration).  Replotting  Fig.~\ref{figapsi100} at different $T$ (see also Fig.~\ref{figapsiProb}) reveals that they form travelling waves, with a peak  moving in towards $a^2=a^2_\star=k\phi_0$ for $T<0$, then  moving  out for $T>0$, after a reflection at $a^2=a^2_\star$ and $T\sim 0$. This happens without the need to appeal to Vilenkin's travelling waves (for $\Lambda>0$). Hence the ``outgoing wave'' of Vilenkin is realized by HH states within unimodular theory for large positive times. It will be analytically proved in the next subsection that at large $|T|$ the peak of the packets follows the classical trajectory.
   %
    \item The only time when the wave function is real is at $T=0$, when we have a superposition of perfectly symmetric incident and a reflected travelling waves. This forms an instantaneous standing wave, depicted in Fig.~\ref{figapsiT0}.
    (For $|T|<\sigma_T$ there is interference between the two waves, even if they are not symmetric.)
    In the bottom panel of  Fig.~\ref{figapsiT0} (a close up of the top panel around $a^2_\star$) we can also seen the evanescent wave penetrating the classically forbidden region ($a^2<a_\star ^2$) and smoothly extending to the Euclidean region at  $a^2<0$. There is nothing special about the point $a^2=0$ corresponding to the ``no-boundary proposal''. 
    
 \begin{widetext}
  \end{widetext}  
    \begin{figure}
\centering
\begin{tabular}{c c}
    \includegraphics[scale=0.53]{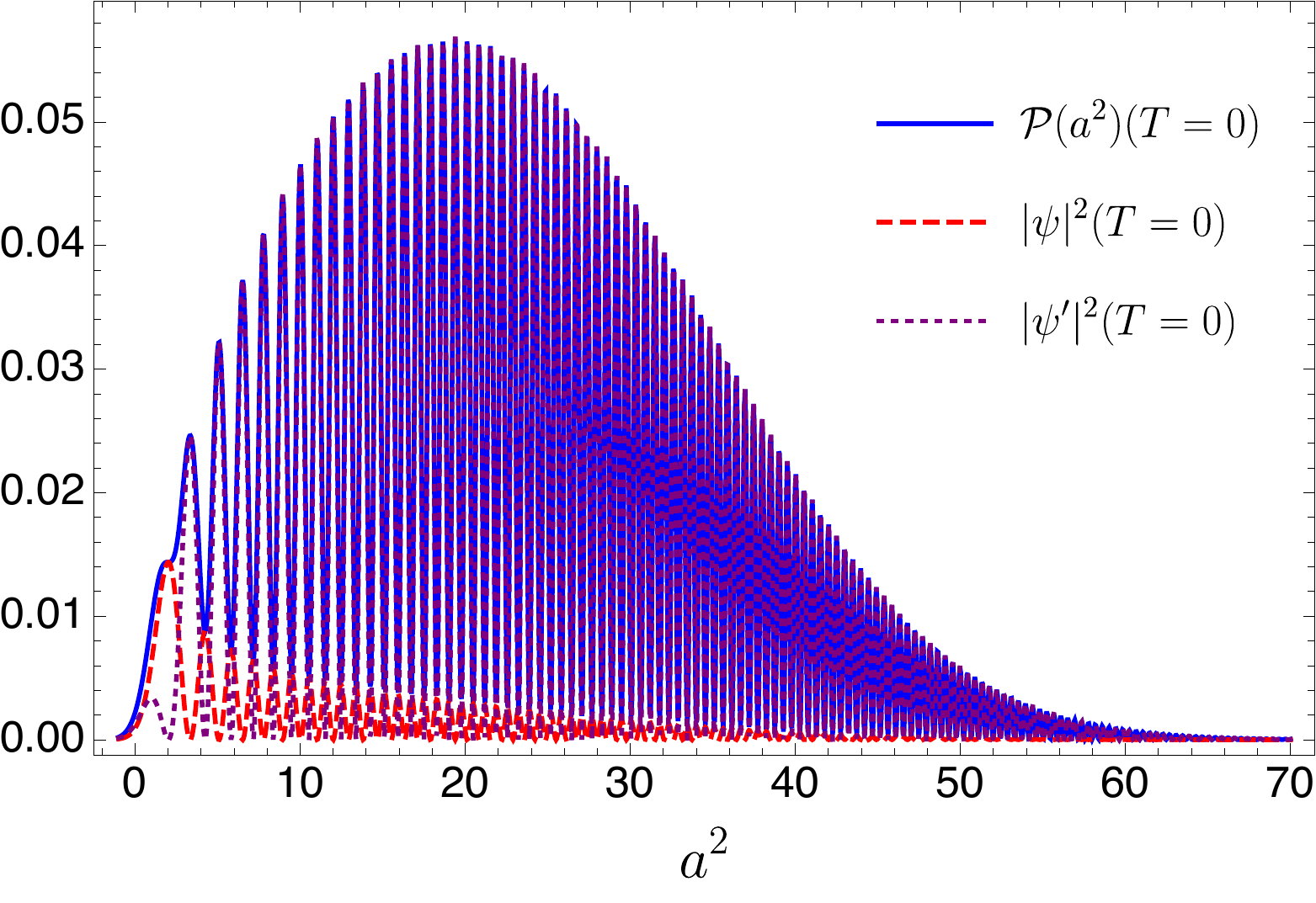} & \includegraphics[scale=0.53]{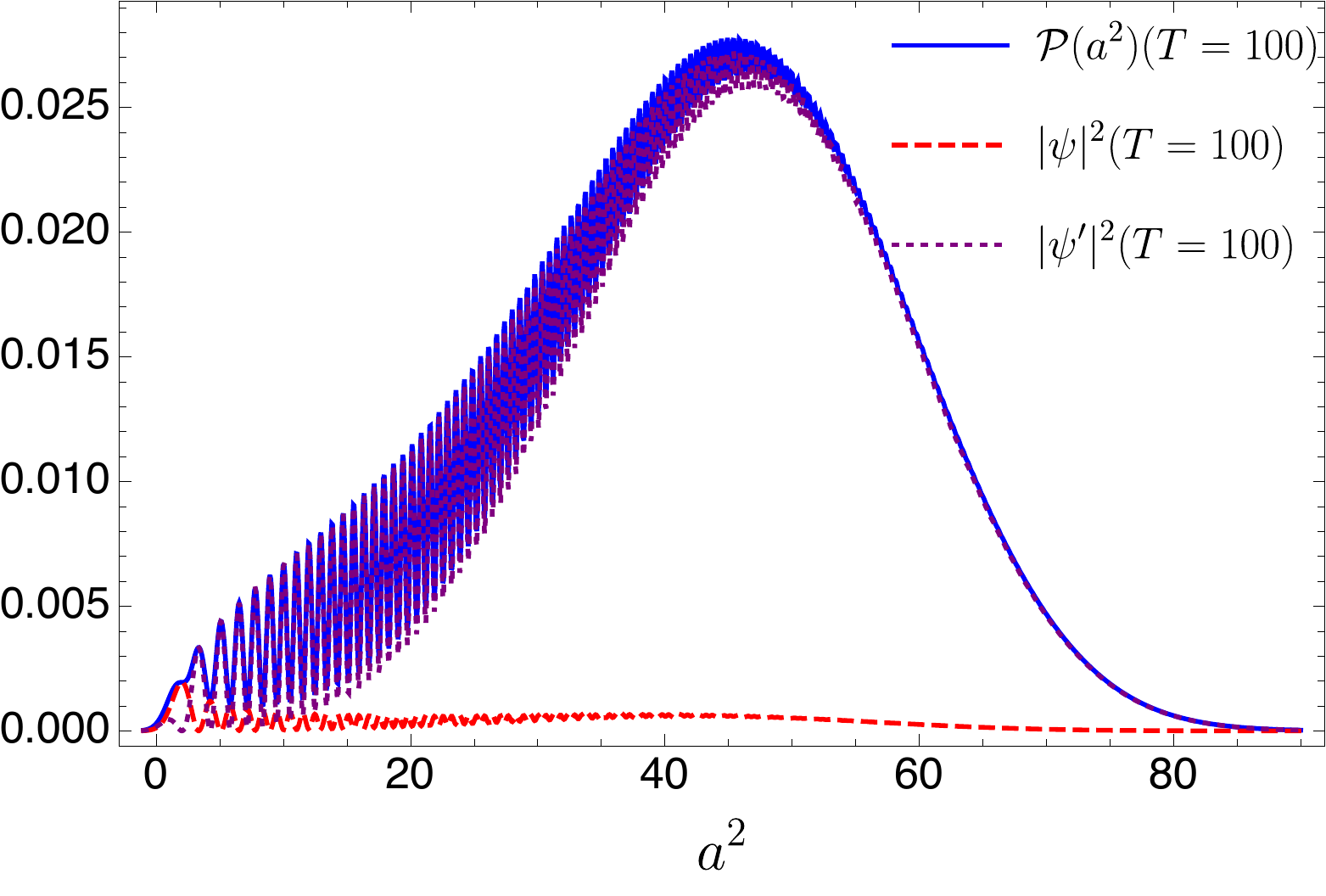} \\
\includegraphics[scale=0.53]{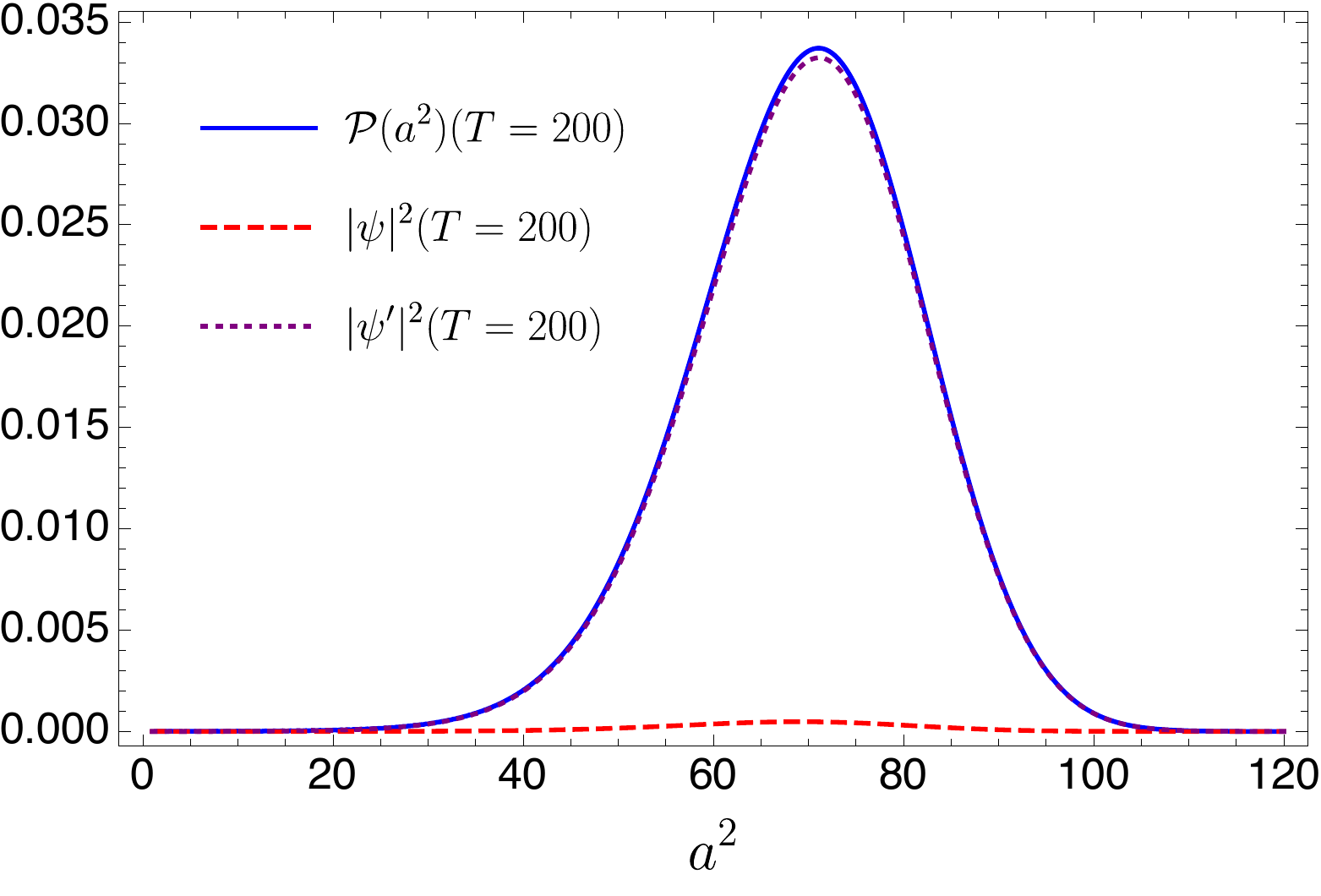}  & 
\includegraphics[scale=0.53]{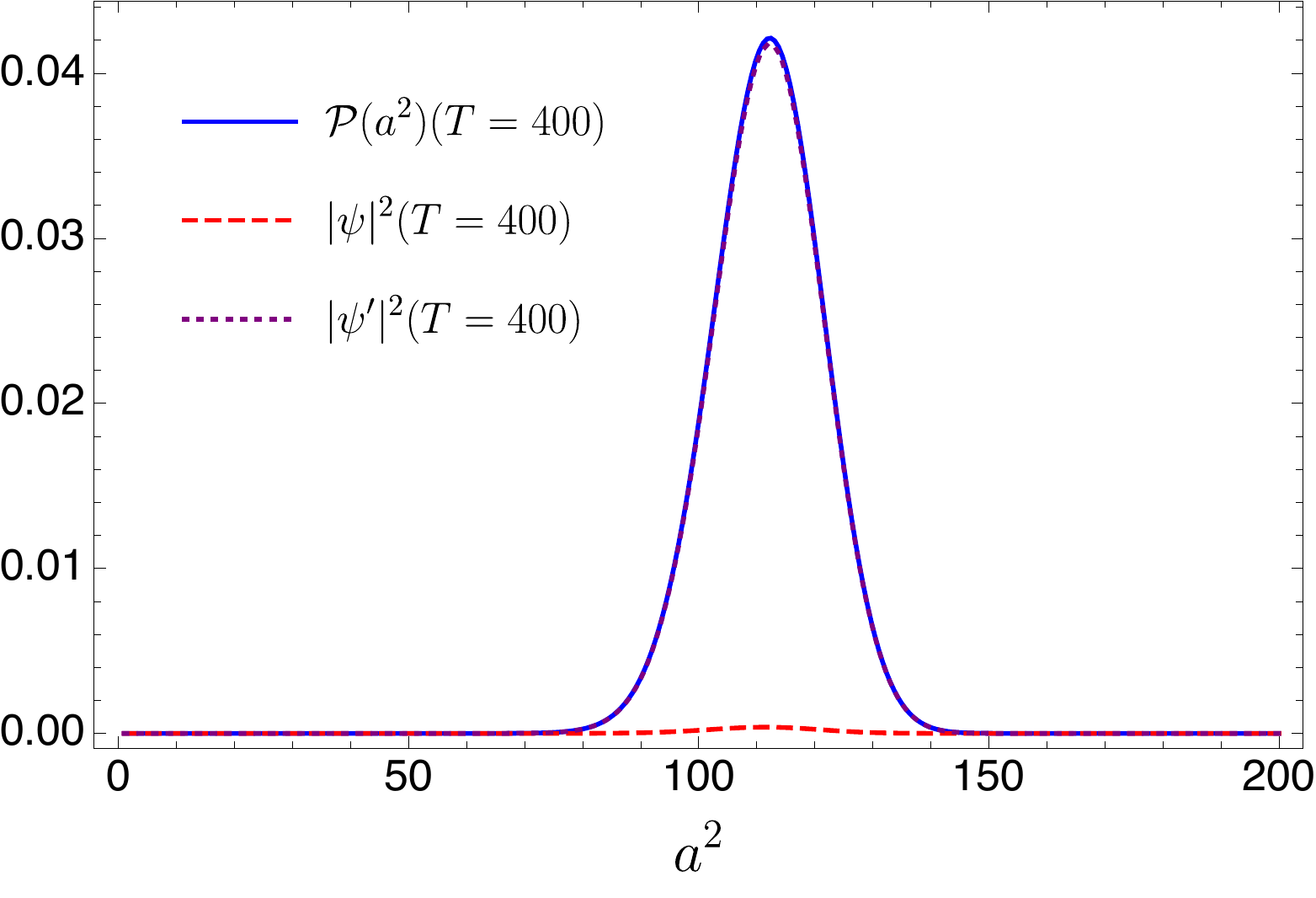}
\end{tabular}
\caption{Probability 
${\cal P}(a^2)$, according to Eq.~(\ref{proba2}), for times $T=0,$~$100,$~$200$ and 400 (still with the same parameters, which we recall imply $\sigma_T=50$).
For $|T|\sim \sigma_T$ the probability has many oscillations inside the envelope, but these disappear at $|T|$ increases, as illustrated. We have also plotted the Born and gradient contributions to the probability according to (\ref{proba2}). The gradient contribution dominates in the semiclassical limit. We see that for increasing $|T|\gg \sigma_T$, the width of the probability {\it decreases}, that is the uncertainty in $a$ decreases.} 
\label{figapsiProb}
\end{figure}

    \item These features are confirmed by an evaluation of the probability  
${\cal P}(a^2)$ defined by Eq.~(\ref{proba2}), as depicted in Fig.~\ref{figapsiProb}. Indeed, the probability travels in tandem with the envelope of the wave packet.
For $|T|< \sigma_T$ the probability has many oscillations inside the envelope, but these disappear at $|T|$ increases, as illustrated.  Furthermore, the width of the probability {\it decreases}, i.e. the packets get {\it sharper}, as $|T|$ increases. This is in contrast with other metric formulations~\cite{bbounce} and will be proved analytically in the next subsection. 

\item 
In Fig.~\ref{figapsiProb}
we have also plotted the contribution to ${\cal P}(a^2)$ of the two terms contributing to (\ref{proba2}): the Born-like term $k|\psi|^2$ and the gradient term $\plk^2|\partial_{a^2}\psi|^2$. As already announced around Eq.~\ref{proba2approx}, the gradient term is dominant in the semiclassical limit. In fact, 
the only occasion when the Born term, $|\psi|^2$, dominates is around $T=0$ (or for $|T|<\sigma_T$) and for $a^2<a_\star ^2$ (or for $a^2>a_\star ^2$ but $a^2\sim a_\star ^2$). For an illustration see 
Fig.~\ref{figapsiProbT0}. 
    \begin{figure}
\centering
\includegraphics[scale=0.53]{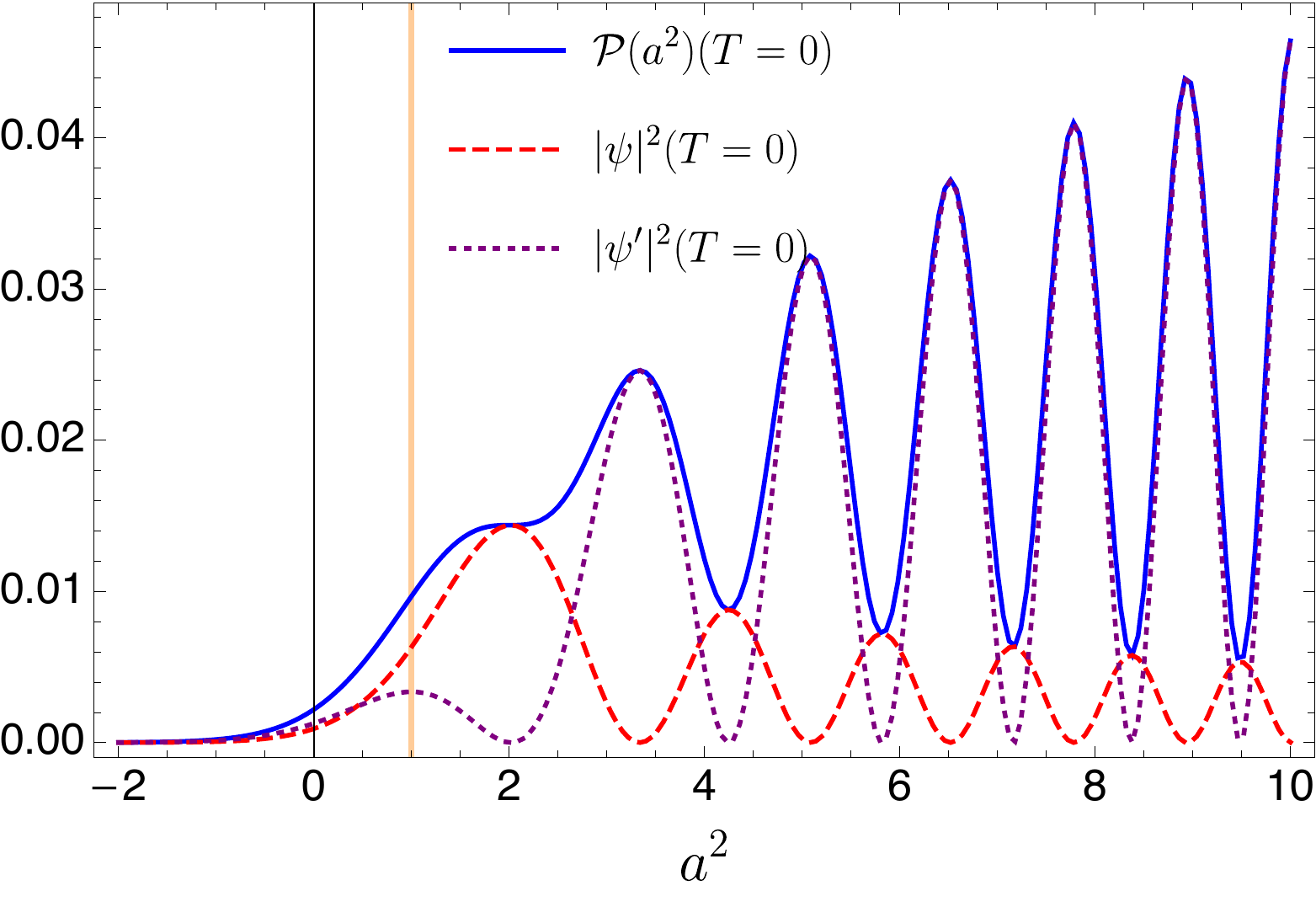}
\caption{Close up of the probability in metric space as a function of $a^2$ at $T=0$, and the respective contributions of the module squared of the wave function and its derivative. The only occasion when the Born term, $|\psi|^2$, dominates is around $T=0$ (or for $|T|<\sigma_T$) and for $a^2<a_\star ^2$ (or for $a^2\gtrsim a_\star ^2$). 
The orange vertical line represents $a=a_*.$} 
\label{figapsiProbT0}
\end{figure}


\end{itemize}

\subsection{Semiclassical limit}

The WKB approximate solutions are useful for understanding the behavior of our analytically exact (but numerically obtained) solutions, as well as for proving that for large $|T|$ the packets follow the classical trajectory. 

We start with the $\psi_s$. 
Well away from $a_\star^2$
($z=0$)  we can use the approximations for the Airy function:
\begin{eqnarray}
{\rm Ai}(-z)&\approx& \frac{1}{\sqrt{\pi}z^{1/4}}\sin\left(\frac{2}{3}z^{3/2}+\frac{\pi}{4}\right)\label{aiapprox1}\\
&\approx& \frac{1}{2\sqrt{\pi}|z|^{1/4}}\exp\left(-\frac{2}{3}|z|^{3/2}\right)\label{aiapprox2}
\end{eqnarray}
valid for $z\gg 1 $ and $z\ll 1$ respectively (the last case included for later reference). Assuming $|T|\gg \sigma_T$ (semiclassical regime), in the construction of our wave packets we will only need the regime where $a^2\gg a_\star^2$, and so:
\begin{eqnarray}
z\approx  \frac{a^2}{\phi^{1/3}\plk^{2/3}}\gg 1
\end{eqnarray}
leading to:
\begin{eqnarray}
\psi_s(a^2,\phi)&\approx&
\frac{C(\phi)}{\sqrt{a}} \sin \left(\frac{2}{3} \frac{a^{3}}{\sqrt{\phi} \plk}
\right)
\label{eqpsiwkb}
\end{eqnarray}
with:
\begin{eqnarray}
C(\phi)&=& \frac{1}{\sqrt{\pi\plk}\phi^{1 / 4}} ,
\end{eqnarray}
where we have ignored the $k$ and the $\pi/4$ phases, since these will not matter for wave packet's peak position (even though they do matter for the phases of the beatings within them). We can expand the sine wave into two complex terms:
\begin{equation}
    \sin \frac{P}{\plk}=\frac{e^{\frac{i}{\plk}P} - 
    e^{-\frac{i}{\plk}P}
    }{2i }
\end{equation}
with wave number:
\begin{eqnarray}
P(a,\phi)= \frac{2}{3} \frac{a^{3}}{\sqrt{\phi}},
\end{eqnarray}
so that inserting the approximate $\psi_s$ into the integral (\ref{HHpacketsa2}) converts the timeless HH standing wave into superposition of two identical travelling waves moving in opposite directions:
\begin{eqnarray}
\psi(a^2,T)&=&\int d\phi
{\cal A}(\phi)\frac{C(\phi)}{2i\sqrt{a}}
(e^{
-\frac{i}{\plk}(\phi T-P)}-e^{
-\frac{i}{\plk}(\phi T + P)}),\nn
\end{eqnarray}
as a result of the time complex phase $\sim \exp(-i\phi T/\plk)$. For a delta function ${\cal A}(\phi)$ (and so $\sigma_T=\infty$) this leads to a standing wave: all that changed was the insertion of a time factor in an otherwise timeless wave function (indeed, a standing wave can be seen as a superposition of identical travelling waves moving in opposite directions). 

For a Gaussian (\ref{Gaussamp}) amplitude, however, the story is very different. 
Taylor expanding $P$ around $\phi=\phi_0$ (mimicking the procedure for the connection representation; see~\cite{JoaoPaper}), and keeping all slow-varying factors at their peak value (since we are not interested in the overall normalization, but only the peak position), we find:
\begin{equation}
   \psi(a^2,T)=\frac{C(\phi_0)}{2i \sqrt{a} (2\pi\sigma_T^2)^{1/4}}(\psi_+(a^2,T)+\psi_-(a^2,T))\\
\end{equation}
with incident ($+$) and reflected ($-$) wave packets:
\begin{equation}
     \psi_\pm (a^2,T)=e^{
     -\frac{i}{\plk}(\phi_0 T \mp P(a^2,\phi_0))}e^{-\frac{(X_{\text{eff}}\mp T)^2}{4\sigma_T^2}}, 
\end{equation}
where:
\begin{eqnarray}
X_{\text{eff}}=\frac{\partial P}{\partial\phi}\Bigg|_{\phi_0}=-\frac{a^3}{3\phi_0^{3/2}}
\end{eqnarray}
and $\sigma_T=\plk/(2\sigma_\phi)$ saturates 
the Heisenberg uncertainty relation following from (\ref{com2}), just as was found (with fewer approximations) in the connection representation (cf. Eqs.~(\ref{Gausspacketsb}) and~(\ref{sigmaT})). 

Hence, the wave packets are localized to within $\sigma(X_{\rm eff})\sim \sigma_T$, and given that the sign of $X_{\rm eff}$ is fixed, but the sign of $T$ changes for the two waves, only one of them is unsuppressed for $|T|\gg \sigma_T$ (the $+$ wave for $T<0$ and the $-$ wave for $T>0$). Consequently they do not interfere in this regime, forming independent travelling waves, dominating at different epochs (signs for $T$). Moreover, since their peaks are at  ${X}_{\text{eff}}=\pm {T}$, they follow the classical trajectory for a contracting an expanding de Sitter Universe, respectively, given that
\begin{eqnarray}
&&\dot{T}=\frac{a^3}{\phi_0^2} \\
&&\dot{X}_{\text{eff}}=-\dot{a}\frac{a^2}{\phi_0^{3/2}}
\end{eqnarray}
reproduces the Friedmann equation:
\begin{eqnarray}
\frac{\dot{a}}{a}=\mp\frac{1}{\sqrt\phi_0}=\mp\sqrt{\frac{\Lambda_0}{3}}
\end{eqnarray}
for the incident  and  reflected waves.
Near the reflection at $T=0$, however, the incident and reflected waves are superposed and interfere. For $|T|\ll \sigma_T$ we recover a standing wave dressed by an exponential fall off, as illustrated in Figs.~\ref{figapsiT0},~\ref{figapsiProb} and~\ref{figapsiProbT0}.

The surprising thing here is that the packets get sharper as $|T|$ increases. This can be derived using error propagation and $\sigma (X_{\rm eff})=\sigma_T$, because:
\begin{equation}
    \frac{\sigma(a)}{a}\approx \frac{1}{3} \frac{\sigma(X_{\rm eff})}{X_{\rm eff}}\approx  \frac{\sigma_T \phi_0^{3/2}}{a^3}.
\end{equation}
The phenomenon is illustrated in Fig.~\ref{figapsiProb}. 
What is surprising is that this happens in tandem with the peaks in $b$ sharpening up as shown, e.g. in~\cite{JoaoPaper}, but can also be seen from (\ref{Gausspacketsb}) using the same method:
\begin{equation}
     \frac{\sigma(b)}{b}\approx \frac{1}{3} \frac{\sigma(X_{\rm CS})}{X_{\rm CS}}\approx  \frac{\sigma_T}{b^3}.
\end{equation}
Hence, for this theory, 
as the Universe gets larger (or when it was larger, before the bounce) the packets in both $b$ and $a^2$ are sharper, in apparent contradiction with the naive Heisenberg uncertainty relation following from (\ref{com1}) (see also \cite{barrow-heis}).


\subsection{Non-hermiticity of $a^2$ operator}
 
 This apparent contradiction with the Heisenberg principle has an explanation. Hermiticity is an essential assumption in the derivation of the Heisenberg-Schr\"odinger principle. Starting from the connection representation we are led to the inner product (\ref{innX}), in terms of which $b$ and the Hamiltonian are Hermitian; however $a^2$ is not.  
 
 This follows directly from examining whether the identity:
    \begin{equation}
        \langle\psi_1|\hat {\cal O}\psi_2  \rangle-\langle\psi_2|\hat{ \cal O}\psi_1  \rangle^\star=0
    \end{equation}
    is valid for generic normalizable states. Using Eq.~(\ref{innX}) we find that this is true for ${\cal O}=\hat b$ as long as $b$ is real (as is the assumption in this paper).
    For ${\cal O}=a^2$ we have:
    \begin{equation}\label{hermit1}
        \langle\psi_1|\hat a^2 \psi_2  \rangle-\langle\psi_2| \hat a^2 \psi_1  \rangle^\star=2 i\plk \int db\, b \psi_1^\star \psi_2,
    \end{equation}
    since the required integration by parts hits the $b$ dependent factor in the measure $dX=db(b^2+k)$. 
    The RHS of (\ref{hermit1}) becomes  more and more prominent the larger the $b$ where the wave packet peaks, and so the more classical the regime. This explains why the Heisenberg principle is violated precvisely in the semiclassical regime.

    Note that the same conclusion could be drawn by evaluating the inner products in (\ref{hermit1}) in the $a^2$ representation, using Eq.~(\ref{inna2}). Performing this exercise we see that the non-hermiticity of $a^2$ now arises from the second term in (\ref{proba2}), that is the gradient term. This is consistent with the observation that this term dominates in the classical regime.

    The non-hermiticity of $a^2$ is not a problem. We stress that its eigenvalues are real. There is extensive literature claiming that Hermiticity is overrated in quantum mechanics, and that one should relax the criterion for what constitutes observables and acceptable Hamiltonians. Real eigenvalues seem to be more important than Hermiticity~\cite{POVM-review,bender}\footnote{We stress that our Hamiltonian is Hermitian.}. Ironically, some of this work arises from attempts to  define time (in standard quantum mechanics) perhaps more conventional than the unimodular proposal (in quantum cosmology) presented here.

    As an object lesson, this example demonstrates how the Heisenberg relations may be more dynamical than they seem.  True, the Heisenberg-Schr\"odinger derivation of the uncertainty relations only relies on the commutation relations and hermiticity, resulting from a straightforward application of the Cauchy-Schwarz inequality for an assumed inner product. The latter is trivial in the standard theory, and has no dynamical input. But in non-standard situations, such as some approaches to quantum gravity, the inner product can have a strong dynamical input (for example the constraints may be implemented as a condition on physical states via the inner product, rather than as operator conditions). In our case the inner product was directly suggested by the dynamics in the connection representation, shaping the violation of one of the assumptions of the Heisenberg principle (hermiticity), since this violation of hermiticity happens with respect to this inner product.

\section{The unimodular propagators}\label{unimodpropagators}
We now proceed to make first contact with the ideas of quantum creation from nothing by evaluating the unimodular propagators and showing how they relate to those in the fixed-$\Lambda$ theory. Elsewhere we will evaluate these propagators using the path integral formalism~\cite{RayPaper}. Here we show how they can be read off directly from the formalism developed so far.

The unimodular propagators are defined from:
\begin{equation}\label{Greendef}
    \psi(q,T)=\int d\mu(q') G(q,T;q',T')\psi(q',T').
\end{equation}
with suitable measure $d\mu(q')$, where $q=(b,a^2)$ and $q'=(b',a^{2\prime})$ (that is, we have connection, metric and mixed propagators). 
They can be read off from the general expansions in terms of the amplitudes evaluated at time $T$, together with the expressions of the amplitudes in terms of the wave function evaluated at time $T'$. 

For $q=b$ and $q'=b$, choosing $d\mu=dX(b')$,  
Eqs.~(\ref{HHpackets}) and (\ref{Ampb}) lead to:
\begin{eqnarray}
 G(b,T;b',T')&=&\int d\phi\, \psi_\phi(b,T)\psi^\star_\phi (b',T')
 \nn\\
 &=&\int d\phi\, e^{-\frac{i}{\plk } \phi \Delta T} \psi_s(b,\phi)\psi_s^\star(b',\phi) 
 \nn\\
 &=&
 \delta(X-X'-\Delta T),\label{Greenb}
\end{eqnarray}
with $\Delta T=T-T'$. 
As already pointed out around (\ref{FTXalpha}), 
this could have been guessed from the fact that the waves are non-dispersive in $X$ (or that $X$ is a linearizing variable in MSS~\cite{JoaoPaper,DSR}), so that any solution must be a function of $X-T$.  
Hence the solution at any time takes the same functional form once we retard it by  $\Delta T$, explaining the delta function in the propagator.  


For $q=a^2$, choosing $d\mu(a^2)=da^2 a^2/\phi$ in (\ref{Greendef}), using (\ref{HHpacketsa2}) and (\ref{ampa2b}) we recover 
 (\ref{Greendef}) with propagator:
\begin{eqnarray}
 G(a^2,T;a^{2\prime},T')&=&\int d\phi\, \psi_\phi(a^2,T) \psi_\phi ^\star(a^{2\prime},T')
 \nn\\
 &=&\int d\phi\, e^{-\frac{i}{\plk } \phi \Delta T} \psi_s(a^2,\phi)\psi_s^\star(a^{2\prime},\phi)
\label{Greena2}
\end{eqnarray}
where we stress that the $\psi_s$ are only real Airy functions if $\phi>0$ (cf. (\ref{psisAi}) and (\ref{psisaibi}); this will be important in some manipulations). 

Likewise we can derive similar expressions for the mixed propagators, adjusting the integration measure $d\mu(q')$ to whatever variable is chosen for $T'$.


As before, these expressions can be seen as insertions of completeness relations.
Eq.~(\ref{Greendef}) is an insertion of 
\bea
\mathbb{1}=\int d\mu(q') \, |q' T'\rangle  \langle q',T' |
\eea
into $\psi(q,T)=\langle q T| \psi\rangle$, with:
\begin{equation}
    G(q,T;q,T')\equiv \langle q,T |q',T'\rangle,
\end{equation}
assuming a further insertion of Eq.(\ref{1phi}), giving:
\begin{eqnarray}
   G=\langle q,T |q',T'\rangle &=&
   \int d\phi \, \langle q,T|\phi\rangle\langle \phi |q',T'\rangle\nn\\
&=&\int  d\phi \, \psi_\phi   (q,T) \psi^\star_\phi  (q',T'),\label{Greengen} \end{eqnarray}
that is, we recover (\ref{Greenb}) and (\ref{Greena2}).
  Moreover, we can relate the propagators in unimodular theory with those in the fixed-$\Lambda$ case. Applying (\ref{factorbracket}) to (\ref{Greengen}) we have:
\begin{eqnarray}
\langle q,T |q',T'\rangle&=&
   \int d\phi\, e^{-\frac{i}{\plk } \phi \Delta T} \langle q|\phi\rangle\langle \phi |q' \rangle
\end{eqnarray}
so that, noting that the  fixed-$\Lambda$ propagators are\footnote{This is consistent with the path integral literature~\cite{Jonathan} if the correct contours are chosen (even when wave functions in terms of ``the momenta conjugate to $a$'' are not recognized as the CSK state).}:
\begin{equation}
    \langle q|q' \rangle_\phi=\langle q|\phi\rangle\langle \phi |q' \rangle= \psi_s(q;\phi) \psi_s^\star (q';\phi),
\end{equation}
we have:
\begin{eqnarray}\label{propsunimodrelfix}
\langle q,T |q',T'\rangle&=&
   \int d\phi\, e^{-\frac{i}{\plk } \phi \Delta T} \langle q|q' \rangle_\phi .
\end{eqnarray}
Hence, the unimodular propagators are related to, but should not be confused with the fixed-$\Lambda$ propagators, even setting $\Delta T=0$.


\subsection{Another form for the propagators}

``Retardation'' is a useful concept in the connection representation, since (\ref{Greenb}) can be written as:
\begin{equation}
    \langle b T|b' T'   \rangle=\delta(X'-X_{\rm ret})=\frac{\delta(b'-b_{\rm ret})}{b'^2+k}.
\end{equation}
with:
\begin{equation}
    X_{{\rm ret}}(b,\Delta T)\equiv X(b)-\Delta T
\end{equation}
defining $b_{\rm ret}=b(X_{\rm ret})$. Solving for the only real solutions of the cubic equation, we obtain explicitly:
\begin{equation}
 b_{\rm ret}=\frac{\sqrt[3]{\sqrt{4 k^3+9 X_{\rm ret}^2}+3 X_{\rm ret}}}{\sqrt[3]{2}}-\frac{\sqrt[3]{2}
   k}{\sqrt[3]{\sqrt{4 k^3+9 X_{\rm ret}^2}+3 X_{\rm ret}}}   \nn 
\end{equation}
(which simplifies to $b_{\rm ret}=\sqrt[3]{b^3-3\Delta T}$ if the curvature $k$ can be neglected). 

We can evaluate the other propagators inserting the integral representation of the Airy functions into (\ref{propsunimodrelfix})\footnote{The subtlety pointed out around Eq.~(\ref{psisaibi}) is relevant here.}. We find for the mixed propagator:
\begin{eqnarray}
\langle b T|a^2  T' \rangle&=&\frac{e^{\frac{i}{\plk }b_{\rm ret} a^2}} {\sqrt{2\pi \plk}(b_{\rm ret}^2+k)},\\
\langle a^2 T| b  T' \rangle&=&\frac{e^{-\frac{i}{\plk }b_{\rm adv} a^2}} {\sqrt{2\pi \plk}(b_{\rm adv}^2+k)},
\end{eqnarray}
(with $X_{{\rm adv}}(b,\Delta T)\equiv X(b)+\Delta T$ and $\Delta T$ still defined as  $\Delta T=T-T'$).
For the metric propagator we have:
\begin{eqnarray}
 \langle a^2 T|a^{2\prime} T' \rangle
 &=&\int \frac{db}{2\pi\plk}\frac{e^{-\frac{i}{\plk}(b a^2-b_{ret}a^{2\prime})}}{b_{\rm ret}^2+k}\nn\\
 &=&\int 
 \frac{db_{\rm ret}}{2\pi\plk}
 \frac{e^{-\frac{i}{\plk}(b a^2-b_{ret}a^{2\prime})}}{b^2+k}
\end{eqnarray}
(since $db_{\rm ret}(b^2_{ret}+k)=
db(b^2+k)$). 
In some cases this simplifies, for example the equal-time expression is:
\begin{equation}\label{eqtpropa2}
    \langle a^2 0|a^{2\prime} 0 \rangle
    =\frac{e^{\frac{-|\Delta a^2|\sqrt{k}}{\plk}}}{2\sqrt{k}\plk}.
\end{equation}
 
\subsection{The equal-time norms of the metric and connection states}
From the above we collect the equal-time limit of the propagators:
\begin{eqnarray}
\langle b 0|b'  0\rangle&=&\delta(X'-X)=\frac{\delta(b'-b)}{b'^2+k},\label{eqtpropb}\\
\langle b 0 |a^2  0 \rangle&=&\frac{e^{\frac{i}{\plk }b a^2}} {\sqrt{2\pi \plk}(b^2+k)}\\
\langle a^2 0|a^{2\prime} 0 \rangle&=&\frac{e^{\frac{-|\Delta a^2|\sqrt{k}}{\plk}}}{2\sqrt{k}\plk}\label{eqtpropa}. 
\end{eqnarray}
We see that  the $b$ eigenstates are orthogonal (in fact, orthonormal with respect to measure $dX$), and that the $a^2$ and $b$ eigenstates
have the inner product expected from the fact that they are duals (still accounting for the $dX$ measure). However, the  $a^2$ eigenstates are 
not orthogonal, their inner product displaying an exponential fall off with correlation length of the order of $\sqrt {\plk} k^{1/4}$.  This implies an intrinsic smearing of this order in any measurement in $a^2$ (like in some on POVMs~\cite{POVM-review}).  

That the eigenvectors of $\hat a^2$ are not orthogonal is to be expected and  explains how its eigenvalues can be real whilst the operator is non-Hermitian. In the traditional quantum mechanics literature the suggestion is often made that if an operator has real eigenvalues but is not Hermitian, then the inner product could be redefined so that it does become so. But that cannot be done for both metric and connection, at least within the framework proposed here.

\section{Retailoring  propagators for unimodular theory}\label{retailor}

Creation ``out of nothing'' and the no-boundary proposal in the fixed-$\Lambda$ metric theory 
involves the propagator: 
\begin{eqnarray}\label{ampvacstar}
 \langle  a^2_\star|0\rangle_\phi&=&{\cal N}_a^2{\rm Ai}(0) {\rm Ai}(-z_0)\nn\\
 &\propto &\exp{\left[-\frac{2}{3} \frac{\phi}{\plk}k^{3/2}\right]}\nn\\
& =& \exp{\left[-\frac{6 V_c k^{3/2}}{l_P^2\Lambda}\right]}
\end{eqnarray}
(where $z_0\equiv z(a^2=0)=-(\phi/\plk)^{2/3}k$ and approximation (\ref{aiapprox2}) is used in the first step). This is loosely identified with the probability of nucleation of the smallest possible classical Universe (with $a^2=a_\star^2)$ out of nothing (with $a^2=0$). 
One may be tempted to generalize this propagator to unimodular theory as:
\begin{equation}\label{obviousprop}
     \langle a_\star^2 T|0 0 \rangle
=\int \frac{db}{2\pi\plk}
\frac{e^{-\frac{i}{\plk} b a_\star^2}}{b_{\rm ret}^2+k}
\end{equation}
(note that for $T=0$, given (\ref{eqtpropa2}), this is essentially (\ref{ampvacstar})). 
However, this makes no sense. 

Firstly, Eq.~(\ref{Greena2}) shows that any transition from $a^2=0$ to any $a^2$ is an unweighted integral over $\phi$:
\begin{eqnarray}
\langle a^2  T|0 0 \rangle
 &=&\int d\phi\, e^{-\frac{i}{\plk } \phi \Delta T} \psi_s(a^2,\phi)\psi_s^\star(0,\phi)\nn.
\end{eqnarray}
But $a^2_\star =k\phi$ depends on $\phi$. Hence, although this is well-defined in $\langle  a^2_\star|0\rangle_\phi$, it cannot appear in the LHS of (\ref{obviousprop}), or outside the integral in $\phi$. We should instead consider a transition from $a^2=0$ to the generic smallest allowed $a^2_{\rm cl}$, defined as:
\begin{equation}\label{firstHHprop}
     \langle a_{\rm cl}^2 T|0 0 \rangle
=\int d\phi\, e^{-\frac{i}{\plk } \phi \Delta T} \psi_s(a^2_\star ,\phi)\psi_s^\star(0,\phi).\nn
\end{equation}
The Airy integral representation then leads to: 
\begin{equation}
      \langle a_{\rm cl}^2 T|0 0 \rangle
      =\int \frac{db}{2\pi\plk}
\frac{1} {(b^3+3kb + 3T)^{2/3}}.
\end{equation}
instead of the RHS of (\ref{obviousprop}).

This could be an interesting result were it not for a second snag. 
The unimodular propagator we have defined
appears as an unweighted integral over $\phi$, and we may not want to assume that $\phi$ is uniformly distributed. But with the standard definition of propagator this is unavoidable. Propagators are defined such that whatever we fix at the end points must leave the complementary variables totally undefined. They are also defined with fully fixed final and initial times. 
But within unimodular theory (or any other relational time theory), the notion that propagators must have fixed final and initial {\it times}  implies that a totally undefined  complementary variable ($\phi$ in this case) at the end points. 
 Since $\phi$ is a constant of motion, it must be totally undefined at all times, and hence the unweighted integral in the expression for our propagators. 

This is very unsatisfactory. In unimodular theory a totally undefined $\phi$ is the exact converse of a fully fixed one, and both are non-physical because non-normalizable. To be able to account for normalizable states, $\psi$, defined by less extreme ${\cal A}(\phi)$, we must therefore relax the requirement that the initial and final times are precisely defined. We propose the definition:
\begin{equation}\label{generalprop}
    { \langle a^2 T|a^{2\prime} 0 \rangle}_\psi
=\int d\phi\, {\cal A}(\phi) e^{-\frac{i}{\plk } \phi T} \psi_s(a^2 ,\phi)\psi_s^\star(a^{2\prime},\phi)\nn
\end{equation}
with an implicit smearing in the initial and final $T$. 
Note how this reduces to 
\begin{equation}
    { \langle a^2 T|a^{2\prime} 0 \rangle}_\phi=
     e^{-\frac{i}{\plk } \phi T} 
    { \langle a^2|a^{2\prime} \rangle}_\phi
\end{equation}
for a (non-normalizable) delta function in $\phi$. This is just the fixed-$\Lambda$ theory's propagator times the monochromatic time evolution factor.

In view of what we said, we therefore argue that a better definition for the creation out of nothing amplitude in unimodular theory is:
\begin{equation}\label{secondHHprop}
     \langle a_{\rm cl}^2 T|0 0 \rangle_\psi
=\int d\phi\, {\cal A}(\phi) e^{-\frac{i}{\plk } \phi \Delta T} \psi_s(a^2_\star ,\phi)\psi_s^\star(0,\phi).\nn
\end{equation}
But even this proves unsatisfactory, since it contains no information about the probability measure of the theory. How would we convert this amplitude into a probability? Would $a^2=0$ be a physically acceptable initial state?

\section{Critique of ``creatio ex nihilo'' in unimodular gravity }
\label{probsnoboundary}


In the standard studies $a=0$ is an isolated point across an effective potential barrier. It makes sense to start the Universe there, either Vilenkin or Hartle-Hawking style. In this Section we examine the problems with this perspective, should we start from a connection based unimodular theory.

\subsection{Problems of $a=0$ as an initial condition in unimodular theory}

In Section~\ref{semiclassical}
we saw how for a Gaussian state in $\phi$ (Eq.~(\ref{Gaussamp})) one can obtain a sound semi-classical limit. Nowhere in the evolution do we find a ``vacuum initial state'', peaked at $a=0$. We can turn the problem around and ask, what amplitude ${\cal A}(\phi)$ would be required for a ``$a^2=0$'' initial state to be a possible solution of the theory? 

Suppose that we define the vacuum initial state as:
\begin{equation}
    \psi(a^2,0)= \frac{\delta(a^2)}{a^2}
\end{equation}
as suggested by the appropriate measure in the propagator (\ref{Greendef}). 
Then (\ref{ampa2b}) implies:
\begin{eqnarray}
{\cal A}(\phi)&=&\frac{{\cal N}_a}{\phi}{\rm Ai}(-z_0)\approx 
\frac{\exp{\left[-\frac{2}{3} \frac{\phi}{\plk}k^{3/2}\right]}}{2\sqrt \pi \phi^{3/2}\plk^{1/2}k^{1/4}}
\end{eqnarray}
for $\phi>0$, and 
\begin{eqnarray}
{\cal A}(\phi)&=&\frac{{\cal N}_a}{2\phi}({\rm Ai}(-z_0)+i {\rm Bi}(-z_0))\approx \\
&\approx&
\frac{i\exp{\left[\frac{2}{3} \frac{|\phi|}{\plk}k^{3/2}\right]}}{2\sqrt \pi |\phi|^{3/2}\plk^{1/2}k^{1/4}}.
\end{eqnarray}
for $\phi<0$. 
Hence, within unimodular theory, the well-known amplitude (\ref{ampvacstar}) is reinterpreted {\it as the amplitude of $\phi$ as implied by a vacuum initial state}. The Universe is then stuck with this probability for $\phi$ forever, which is particularly worrying, as a negative cosmological constant is vastly preferred over a positive one. 

But even before worrying about this, there is another problem: such a state is non-normalizable under (\ref{inna2}). This can also be seen directly from (\ref{innalpha}), or even transforming to the $b$ representation to find $\psi(b)=1$, with the obvious implications under (\ref{innX}). Ditto with:
\begin{equation}
    \psi(a^2,0)\sim \delta(a^2)
\end{equation}
(or any variations thereof). For this state state $\psi(b)=ikb$, equally non-normalizable. 

We can try to evade this, by considering normalizable states (under (\ref{inna2})) which are highly peaked around $a^2=0$: 
\begin{equation}
    \psi(a^2,0)=\frac{\exp^{\frac{-a^4}{4\sigma_a^2}}}{(2\pi \sigma_a^2)^{1/4}}\frac{1}{\sqrt{k+\frac{\plk}{4\sigma_a^2}}}\propto \sqrt{\sigma_a} \exp^{\frac{-a^4}{4\sigma_a^2}},
\end{equation}
where we took the limit $\sigma_a\rightarrow 0$. This suffers from the same problems as the definition of the square-root of a delta function (strictly speaking the wave function is zero). But even ignoring these, we see that it leads to amplitude:
\begin{eqnarray}
{\cal A} (\phi)\sim -\frac{{\cal N}_a\sigma^{7/2}_a}{\phi^{4/3}\plk^{2/3}}{\rm Ai}'(-z_0)\sim \sigma_a^{7/2}
\exp{\left[-\frac{2}{3} \frac{\phi}{\plk}k^{3/2}\right]}\nn
\end{eqnarray}
for $\phi>0$ and a similar expression for $\phi<0$. This has the same asymptotic form as before, with the same problems (a preference for a negative $\Lambda$). It would also imply a Gaussian wave function in $b$, with $\sigma_b=\plk/(2\sigma_a)$, so that:
\begin{equation}
    {\cal P}(b)=(b^2+k)|\psi|^2\sim (b^2+k)\exp\left[-\frac{b^2}{2\sigma_b^2}\right]
\end{equation}
with $\sigma_b\rightarrow\infty$, i.e. a near uniform distribution with two soft peaks at $b=\pm 2\sqrt{2}\sigma_b$, curiously very different from the $b=\pm i$
expected by applying $a^2=0$ to the Hamiltonian constraint.


Note that these issues of measure and normalizability are well-hidden in the standard treatment appealing to propagators.

\subsection{
The unremarkable nature of the nothing}

Furthermore, 
there is nothing special about $a=0$ in the theory we are considering. This follows from the connection representation, even before one adds the unimodular extension. If $b$ is the starting point, then its conjugate is $a^2$ (and not $a$), the densitized inverse metric (see discussion in Section.~\ref{sumary-connection}). The natural range for $b$, and so for its dual $a^2$ as well, is the whole real line. Hence in the dual representation to $b$ we should include the Euclidean section $a^2<0$.

It follows that any creation from nothing theory, as well as the no-boundary proposal, makes little sense here, since $a=0$ is not a ``no-boundary''.  Quite the opposite: it is the fence between Lorentzian and Euclidean spaces. Making it a boundary actually requires suplementary boundary conditions.

We stress that ``Euclidean'' here has a different meaning to that in Hawking's no-boundary proposal. In HH it means  Euclideanizing space right at $0<a^2<a^2_\star$, with a $t=i\tau$ rotation associated the semiclassical treatment of tunneling. Here it means allowing $a^2$ to be negative (keeping time and $N$ real), and no instanton or semiclassical treatment is implied: we evaluated directly the exact evanescent wave. 

In addition, the fact that $a^2<0$ is covered by an evanescent wave implies that in all of our solutions the  $a^2<0$ section is always off-shell, so we should not expect a connection with classical Euclidean GR. On-shell expressions such as $\dot T=Na^3/\phi^2$ or $\dot a=Nb$, valid for the peak of the wave function for semiclassical states, cannot be used there.
Hence there is no contradiction with $b$ being real and $a^2<0$. Also we should not expect $T$ to be imaginary. 


\section{Quantum Creation in unimodular theory}\label{outofwall}

In view of the above it {\it may} make more sense within the quantum unimodular theory we are using to do the following:
\begin{itemize}
    \item Adopt a Gaussian state with $\sigma(\phi)/\phi_0\ll 1$ (or any other sharp function) and use it as a basis for a theory of initial conditions, rather than impose an initial condition in $a$ and endure the implied ${\cal A}(\phi)$ for the rest of the life of the Universe. 
    
    This feature is a consequence of the unimodular extension. 
    
    \item Equate ``quantum creation'' of the Universe with the process by which a semi-classical state at $a^2>a_\star$ emerges from the full region $-\infty<a^2<a^2_\star$, at time $|T|\sim \sigma_T$.  (Note that for the chosen ${\cal A}(\phi)$, $a^2_{\rm cl}$ can be confused with $a^2_\star$ defined by $\phi=\phi_0$.)
    
    This feature is implied by the use of the connection $b$ as the starting point, so that within the metric dual there is nothing unique about $a=0$, $a>0$ or indeed $a^2>0$.  The whole $-\infty<a^2<a^2_\star$ should then be seen as the ``initial state'' of the Universe. 
    
    \item Evaluate the probability for such ``quantum creation'' directly in terms of the wave function and the associated unitary probability, rather than propagators (which may hide issues of normalizability and probability). 
    
    This is a general point. In our case the form of the probability in metric space derives both from the unimodular extension (the probability is originally defined by (\ref{innalpha}) in terms of ${\cal A}$) and the primacy of the connection representation, so that $\phi=3/\Lambda$ is chosen in (\ref{innalpha}), leading to (\ref{innX}) and ultimately (\ref{inna2}). In relation to the previous point, note that the probability only integrates to 1 at all times if the full domain $-\infty<a^2<\infty$ is included. Hence including the Euclidean section in our considerations is non-negotiable once the probability measure is considered, and unitarity in enforced. 
\end{itemize}

We should then backtrack to
Section~\ref{semiclassical}. Therein, we saw that packets of HH wave functions  reproduce the semi-classical limit for all $T$, with the exception of $|T|<\sigma_T$, where the incident and reflected waves interfere, creating a multi-peaked probability (see Fig.\ref{figapsiProb}). Even at $|T|\lesssim\sigma_T$ the bulk of the probability is in $a^2>a^2_\star$: whatever quantum effects are present, they are most relevant in the classically allowed region, with only a small ``penetration'' probability into the wall (see Fig.~\ref{figapsiProbT0}). This small probability, associated with the evanescent wave inside the full infinite wall, should be associated with the probability of quantum creation of the semi-classical Universe.

\begin{figure}
\centering
\includegraphics[scale=0.53]{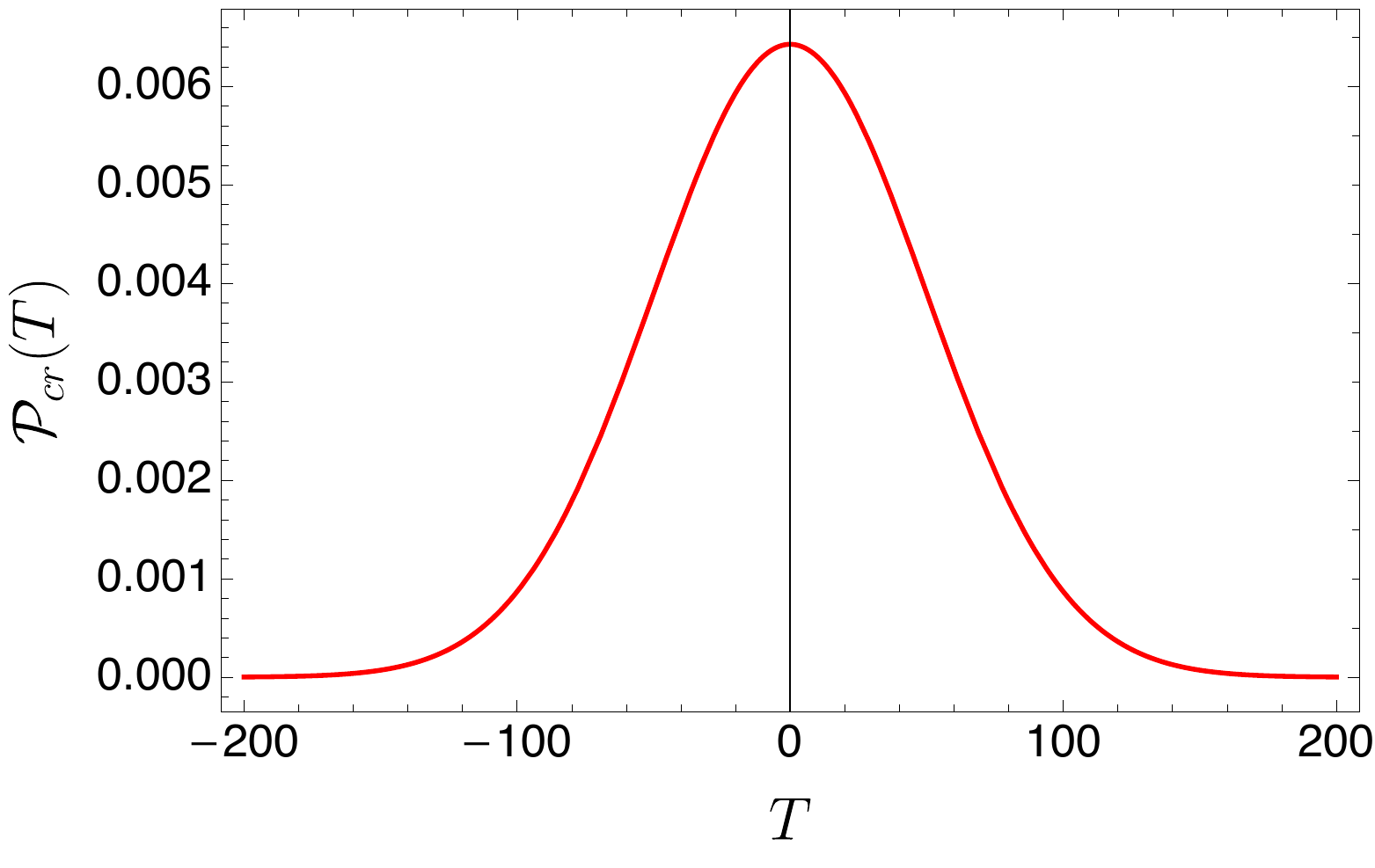}
\caption{The probability of quantum creation of a classical Universe from the classically forbidden region, as a function of $T$ (which cannot be identified with the on-shell expression for unimodular time in this regime). We have used the same illustrative parameters as in Section~\ref{semiclassical}.} 
\label{fig-creation}
\end{figure}
\begin{figure}
\centering
\includegraphics[scale=0.53]{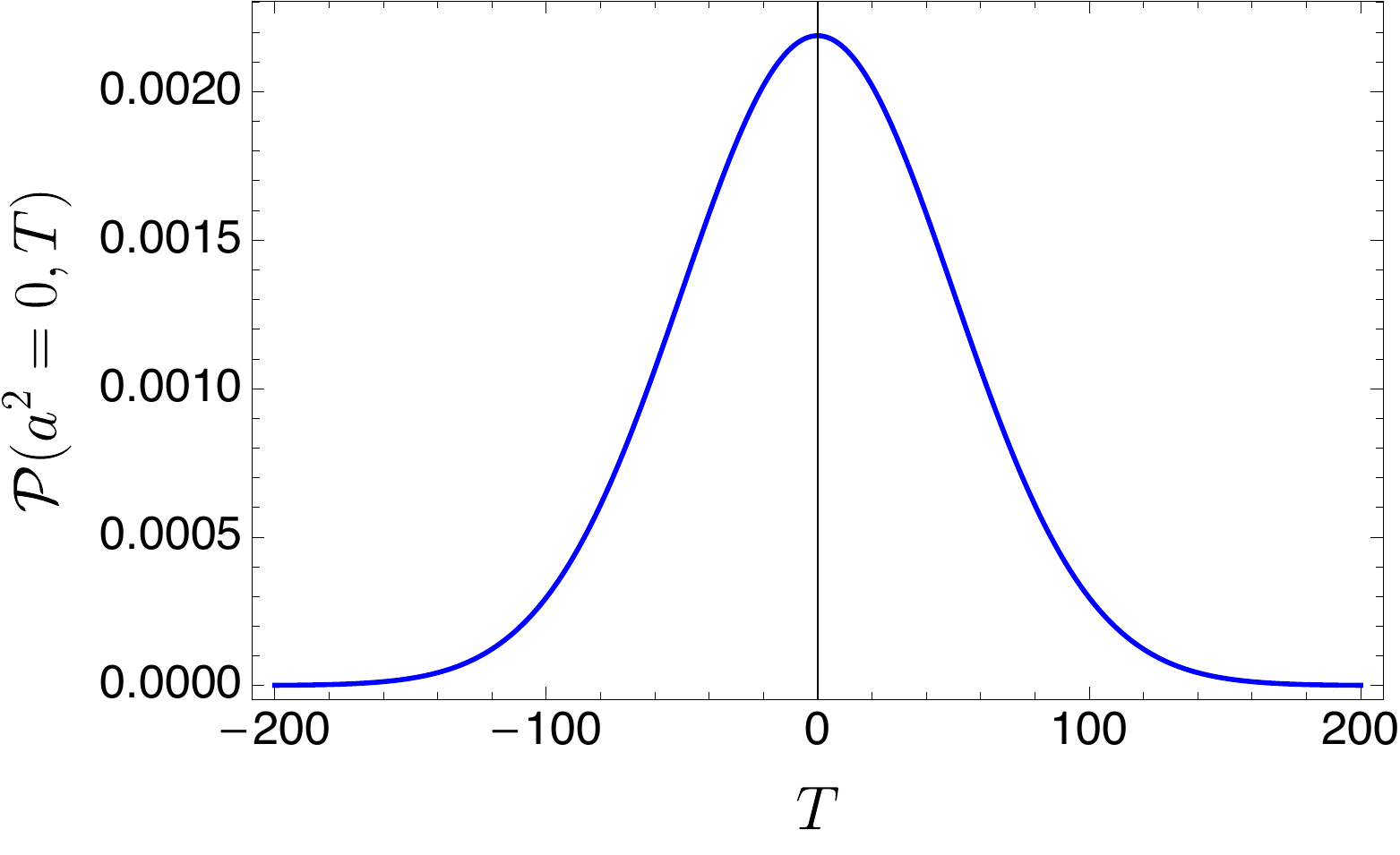}
\caption{Probability ${\cal P}$ at $a^2=0$ as a function of time $T$, given here for comparison.}
\label{fig-probnothing}
\end{figure}

We thus evaluate 
\begin{equation}
    {\cal P}_{\rm cr} (T)=\int _{-\infty}^{a_\star^2(\phi_0)}da^2\; {\cal P}(a^2,T).
\end{equation}
for a Gaussian state (as in Section~\ref{semiclassical}), 
to be interpreted as the probability of the quantum creation of the Universe. 
This probability is a function of time; so, time exists even before the creation of the semiclassical Universe in this scenario. Naturally,  $T$ is a purely quantum variable when $|T|\lesssim\sigma_T$, and can only be identified with unimodular time (i.e. acquire its on-shell expression or ``time formula'', Eq.~(\ref{timeformula})) once the Universe enters its semi-classical regime, for $|T|\gg \sigma_T$. With this proviso, $\sigma_T$ (defined by (\ref{sigmaT})) is the ``time'' it takes to create the Universe in this scenario. We have plotted the numerical result for our illustrative model in Fig.~\ref{fig-creation}. The curve has formal similarities with 
${\cal P}(a^2=0,T)$, depicted in Fig.~\ref{fig-probnothing}, but the physical background is totally different.

How to interpret this process? The fact that we have a Hamiltonian constraint provides leeway not usually found in quantum mechanics. Could the process we have described be seen as the creation of a Universe/anti-Universe pair out of the full classically forbidden regions inside the ``wall region'', including the Euclidean section? The contracting Universe can be seen as the time reversal (the ``anti-particle'') of the expanding semi-classical Universe. The pair can be created without a supply of the energy because the whole system has zero energy throughout. In effect, a pair of time-reversed Universes is spat out of the wall by purely quantum effects. It would be interesting to seek a condensed matter analogue to this process.

\section{Conclusions}
The achievements of this paper are of two types: those regarding the late-life of the Universe and those concerning its birth. 

For the first, a major result is that we do not need the Vilenkin wave function to obtain a late-time outgoing wave function (corresponding to a pure expanding Universe) once an unimodular extension is implemented\footnote{This is not to say that it would not be interesting to investigate unimodular extensions to the Vilenkin wave function. Note the obvious technical hurdle: the fact that $b$ now has a non-real section~\cite{CSHHV} implies that (\ref{FTba2}) and (\ref{FTpack-ab}) would need to be replaced by the Laplace transform. Its inverse would seem to locate the probability on the HH contour only, but there may be subtleties.}. The unimodular Hartle-Hawking wave packets are complex (unlike the real Hartle-Hawking wave function) and already contain  pure outgoing and a pure incoming well-localized and non-interfering waves, with one suppressed and the other dominating in their respective epochs defined by unimodular time ($T>0$ or $T<0$). The two waves only mix around the bounce ($|T|\lesssim \sigma_T$), producing a temporary standing wave. Obviously, letting $\sigma_T\rightarrow\infty$ (i.e. fully fixing $\Lambda$) we recover the standard result, but the picture is very different otherwise. 

Furthermore, 
starting from the connection variable we are led to a natural unitary inner product in the metric representation which is {\it not} the naive Born $|\psi|^2$, and this is most relevant precisely in the semi-classical regime. What is remarkable is that with respect to this inner product $b$ is self-adjoint but  $a^2$ is not, even though its eigenvalues are real. Not only is this non-hermiticity not a problem, but we argue that it is in fact {\it a requirement for a proper semi-classical regime}. If nothing broke down in the assumptions of the Heisenberg principle, then the principle would imply metric-connection squeezing, the peaks in connection space typically becoming sharper and the metric ones wider~\cite{GielenMenendez,bbounce}. It is possible that this would not be observable, but better still is to have a situation where both observables become sharper and sharper as the Universe becomes supposedly classical. Such is the case of the theory in this paper, the tension with the Heisenberg principle resolved by the fact that one of the observables is not self-adjoint.

On the other side of the cosmic story, this paper sheds new light into the possibility that our Universe might have been created as a quantum fluctuation out of nothing. Several problems with this concept have been identified~\cite{JeanLuc1,JeanLuc2}; here we add to the discussion remarks that may be specific to the unimodular extension and connection representation primacy. The`` nothing'' in these theories is non-normalizable and implies a disastrous probability for $\Lambda$. It is also a platitude within the bigger range for the densitized inverse triad implied by its connection dual. If we are going to accept quantum creation, the whole forbidden region, including the Euclidean section (in the sense of $-\infty<a^2<0$,  must be involved for unitarity with respect to the inner product to be preserved. With exact solutions and an inner product in our armory we can then evaluate the probability for a semi-classical Universe to be created. This can be seen as the creation of a pair of Universe/anti-Universes, popping out of the full forbidden wall.  

It would be interesting to see what implications this has for the stability of tensor modes, in particular the damning results of~\cite{JeanLuc1,JeanLuc2} derived  within the path integral metric formalism. What would that matter look like if phrased from the connection starting point, in particular within the canonical formalism and possibly with an unimodular extension? Could the instability identified in~\cite{JeanLuc1,JeanLuc2}
be related to the well-known issues plaguing, or not, the various versions of the Chern-Simons-Kodama state~\cite{witten,lee2,dion,laura,RealCS}? 
And what input into the matter would the unimodular extension offer? We note that, strictly speaking, without this extension (already pioneered in~\cite{UnimodLee1,UnimodLee2})  the Chern-Simons-Kodama state can never be physical or normalizable.

\section{Acknowledgments}
We thank Stephen Gielen, Jonathan Halliwell, Raymond Isichei, Jean-Luc Lehners, David Jennings and Rosinha M. for discussions related to this paper and inspiration. This work was supported by FCT Grant No. 2021.05694.BD (B.A.) and by the STFC Consolidated Grant ST/T000791/1 (J.M.).

\end{document}